\documentclass[showpacs,onecolumn,preprintnumbers,amsmath,amsfonts,amssymb,floatfix,aps,superscriptaddress]{revtex4}

\usepackage{physymb}
\usepackage{graphicx}
\usepackage{epsfig}
\usepackage[hang,nooneline]{subfigure}
\usepackage[normalem]{ulem}
\usepackage{color}
\usepackage{braket}
\usepackage{hyperref}
\usepackage{todonotes}

\newcounter{fig}

\definecolor{darkblue}{rgb}{0.00,0.00,0.55}
\definecolor{black}{rgb}{0.00,0.00,0.00}
\hypersetup{
    linkcolor = darkblue,
    anchorcolor = darkblue,
    citecolor = darkblue,
    filecolor = darkblue,
    urlcolor = darkblue,
    colorlinks  = true,
}

\begin{document}

\title{Computing stationary solutions of the two-dimensional Gross-Pitaevskii equation
  with Deflated continuation}
\author{E. G. Charalampidis}
\email[Email: ]{charalamp@math.umass.edu}
\affiliation{Department of Mathematics and Statistics, University of Massachusetts
Amherst, Amherst, MA 01003-4515, USA}
\author{P. G. Kevrekidis}
\email[Email: ]{kevrekid@math.umass.edu}
\affiliation{Department of Mathematics and Statistics, University of Massachusetts
Amherst, Amherst, MA 01003-4515, USA}
\author{P. E. Farrell}
\email[Email: ]{patrick.farrell@maths.ox.ac.uk}
\affiliation{Mathematical Institute, University of Oxford, Oxford, UK}
\affiliation{Center for Biomedical Computing, Simula Research Laboratory, Oslo, Norway}

\date{\today} 

\begin{abstract}
In this work we employ a recently proposed bifurcation analysis
technique, the deflated continuation algorithm,
to compute
steady-state solitary waveforms in a one-component, two
dimensional nonlinear Schr{\"{o}}dinger equation with a
parabolic trap and
repulsive interactions.  Despite the fact that this system
has been studied extensively, we discover a wide variety of
previously unknown branches of solutions. We analyze the stability of the
newly discovered branches and discuss the bifurcations that
relate them to known solutions both in the near linear
(Cartesian, as well as polar) and in the highly nonlinear regimes.
While deflated continuation is not guaranteed to compute the
full bifurcation diagram, this analysis is a potent
demonstration that the algorithm can discover new
nonlinear states and provide insights into the energy
landscape of complex high-dimensional Hamiltonian dynamical
systems.
\end{abstract}

\maketitle

\section{Introduction} \label{sec:intro}

Over the past two decades, the pristine setting of atomic
Bose-Einstein condensates (BECs) has enabled the exploration
of numerous physical concepts~\cite{book1,book2}. One of the
principal themes examined involves the interface between
nonlinear wave dynamics and such atomic systems, concerning
the study of so-called matter-wave
solitons~\cite{emergent,book_new,dumitru}.  These coherent
structures have not only been theoretically predicted but
also in many cases experimentally observed.  Some
relevant examples include bright~\cite{expb1,expb2,expb3},
dark~\cite{djf} and gap~\cite{gap} matter-wave solitons.
Higher dimensional analogues of these have also been
studied; these permit a wide variety of interesting
structures, such as vortices~\cite{fetter1,fetter2}, solitonic
vortices and vortex rings~\cite{komineas_rev}.

From a bifurcation theoretic perspective, the
one-dimensional repulsive version of this problem is
interesting but not particularly
rich~\cite{alexander,augusto,alfimov} in the customary
experimental setting of a parabolic trapping potential.  One
can systematically trace the bifurcation of nonlinear
branches from the corresponding linear ones. As the
characteristic eigenvalue parameter, the so-called chemical
potential, is detuned from the linear limit of the quantum
harmonic oscillator, nonlinear states emerge via bifurcation
from the trivial branch. However, no subsequent bifurcations
occur on these branches even as the chemical potential
approaches the other analytically tractable limit of large
values, the Thomas--Fermi (TF) limit, where excited states
feature the same profile with embedded single or multiple
dark solitons. 

However, the solutions of the two-dimensional repulsive case
are far more intricate and intriguing.
In this paper we revisit the two-dimensional repulsive
case with a new numerical tool, the deflated continuation
algorithm of Farrell et al.~\cite{farrell1}.
This specific problem is of significant interest for a number
of reasons. It constitutes a prototypical example
where numerous secondary bifurcations
have been shown to occur \cite{middel1,middel2}.
These include symmetry-breaking bifurcations and provide
the potential for genuinely complex order parameter states bearing
vorticity. These bifurcations are also crucial in modifying the
stability characteristics of the states at hand. Hence, as we will
see below, this problem provides a very rich testbed for
investigating the effectiveness of deflated continuation on
an important and intensely studied
problem~\cite{alexander2,middel1,middel2,crasovan,mottonen,virtanen,haque,carr,herring}. From a mathematical perspective,
the analysis of the relevant bifurcations and the resulting
vortex-bearing states are still a topic of recent
investigation~\cite{dmpeli}.

At the heart of deflated continuation is a deflation
technique for computing previously unknown solutions of
differential equations \cite{farrell2}, and we discuss this
first. Suppose we have a nonlinear problem $F(\phi) = 0$,
and one solution $\phi_1$ has been found via Newton
iteration from an initial guess $\bar{\phi}$.  The central
idea of deflation is to construct a new nonlinear problem
$G(\phi) = 0$ with the property that Newton's method is
guaranteed not to converge to $\phi_1$.  Thus, if Newton's
method converges from the same initial guess $\bar{\phi}$,
it will converge to a distinct solution $\phi_2 \neq
\phi_1$. The process can then be repeated until Newton's
method no longer converges to any solution within a fixed
amount of work. In this way, many solutions to the same
problem can be discovered from a single initial guess. However,
there is no guarantee that all of the relevant solutions (for
a given parameter value) will be obtained in this way, as Newton's
method may fail to converge on when applied to $G$ if initialized
far from a solution.

The deflated problem is constructed via the application of a
\emph{deflation operator} to the residual. Suppose $F: U \to V$ is a nonlinear
map between Banach spaces, and $\phi_1$ is an isolated root of $F$, i.e.~
$F'(\phi_1)$ is invertible. In previous work \cite{farrell2}, $G: U \to V$ was constructed via
\begin{equation}
G(\phi) = \left(\frac{1}{\|\phi - \phi_1\|^2_U} + 1   \right) F(\phi),
\end{equation}
where $\|\cdot\|_U$ is the norm on $U$. The essential idea is that
$\|\phi - \phi_1\|^{-2}_U$ approaches infinity as $\phi \to \phi_1$ faster than
$F(\phi)$ approaches 0, and hence
\begin{equation}
\limsup_{\phi \to \phi_1} \|G(\phi)\|_V > 0,
\end{equation}
ensuring that Newton's method will not converge to $\phi_1$ when applied to $G$.

In the present case, this idea must be modified slightly,
as the solutions are no longer isolated: if $\phi_1$ is a solution, then so is its phase
shift $\psi_\theta = e^{i\theta} \phi_1$ for any phase $\theta$. Given knowledge of a solution $\phi_1$,
we therefore wish to deflate the entire group orbit $\{\psi_\theta: \theta \in [0, 2\pi)\}$. This
is achieved by constructing the deflated problem via
\begin{equation}
G(\phi) = \left(\frac{1}{\| |\phi|^2 - |\phi_1|^2\|^2_U} + 1   \right) F(\phi),
\end{equation}
where $|\phi|$ represents the amplitude of the complex-valued wavefunction $\phi$. As the amplitude
is invariant under phase shift, this modified deflation operator eliminates the entire group
orbit, ensuring nonconvergence to any solution trivially related to a known solution. The norm
chosen for $U$ is the $H^1(D; \mathbb{R})$ norm, where $D$ is the domain on which the PDE is posed.

This deflation idea is extended to an algorithm for bifurcation
analysis, called deflated continuation, as follows.
Suppose a set of solutions $S(\mu)$ is known for the nonlinear problem
$F(\phi, \mu) = 0$ at a given
parameter value $\mu \in \mathbb{R}$, and we wish to compute the solutions
at a modified parameter value $S(\mu + \Delta \mu)$. In the
first phase of the algorithm, known branches are continued:
each solution is used as initial guess for the nonlinear
problem at $\mu + \Delta \mu$, and the resulting solution
deflated. In the second phase, new solutions are sought:
each $\phi \in S(\mu)$ is used again as initial guess for
the nonlinear problem at $\mu + \Delta \mu$. In this way, we
allow for the discovery of new branches that have bifurcated
between $\mu$ and $\mu + \Delta \mu$. If a search is
successful, the initial guess $\phi$ is used again. Once
the algorithm has exhausted all initial guesses in $S(\mu)$,
it proceeds to compute $S(\mu +
2\Delta \mu)$ from $S(\mu + \Delta \mu)$, and so on. For a
full description of the algorithm, see~\cite{farrell1}.

There are two central advantages of deflated continuation.
First, the algorithm is capable of discovering
\emph{disconnected} branches, those not continuously
connected to known solutions. Second, the algorithm can scale
to very fine discretizations of PDEs. The only expensive subroutine
required by deflated continuation is the computation of the
Newton step of the undeflated system; thus, if a good
preconditioner is available for the Newton system, the
bifurcation diagram of the system can be efficiently computed.
This will be of central importance in future work on the
corresponding three-dimensional problem (see e.g.~\cite[Chap.~4]{book_new}).

In this problem we have extensive knowledge of the solutions at
the linear limit, as described in section \ref{sec:setup}. We augment
deflated continuation with this knowledge by using these linear solutions
as initial guesses near the bifurcation from the trivial branch. This
augmentation identifies a handful of additional solutions that deflated
continuation alone misses, as will be discussed in section \ref{sec:conclusion}.

We apply deflated continuation to this complex problem
in the hope of gaining new insight into its families of
solutions. We will complement the solutions
identified via deflation with a stability analysis,
aiming at a systematic map of the newly discovered
bifurcations and associated stability changes. In section
II, we will provide a brief overview of the theoretical
setup of both the existence and stability problem.
This will explore 
the analytically tractable linear limit which
will subsequently serve as a way of
potentially seeding nonlinear solutions away from that limit.
In
section III, we will offer a systematic classification of
our numerical results.
Based on a detailed comparison with earlier works
including~\cite{carr,middel2,toddrotating}, we will
provide a wealth of novel families and associated bifurcations.
Finally, in section IV,
we summarize our main findings and discuss the directions of
future study arising from this work.

\section{The model and setup} \label{sec:setup}
We consider the nonlinear
Schr{\"o}dinger (NLS) equation in $(2+1)$ dimensions (two
spatial and one temporal) written in dimensionless form
(see, e.g.,~\cite{emergent,book_new} for details of the
nondimensionalization)
as
\begin{equation}
i\partial_{t} \Phi = -\frac{1}{2} \nabla^{2}\Phi + %
|\Phi|^{2}\Phi + V(\mathbf{r})\Phi.
\label{nls_2D_1_comp}
\end{equation}
Here $\nabla^{2}$ stands for the standard Laplace operator
in 2D and the external potential $V(\mathbf{r})$ assumes the 
standard harmonic form of $V(\mathbf{r})=\frac{1}{2}\Omega^{2}|\mathbf{r}|^{2}$, 
with $|\mathbf{r}|^{2}=x^2+y^2$ and the normalized trap strength
$\Omega$. The latter represents the ratio of trappings along 
and transverse to the plane and should thus be $\Omega \ll 1$; 
in what follows we will fix $\Omega=0.2$. In the context of 
BECs, the (complex) field $\Phi(\mathbf{r},t)$ in Eq.~(\ref{nls_2D_1_comp}) 
represents the macroscopic wave function. The sign of the 
nonlinear term is chosen to reflect the self-repulsive nature
of the interatomic interactions considered herein. 

The construction of stationary solutions is based on the well-known
ansatz
\begin{equation}
\Phi(\mathbf{r},t)=\phi(\mathbf{r})\exp{\left(-i\mu t\right)}, 
\label{stat_ansatz}
\end{equation}
with chemical potential $\mu$. By substituting Eq.~(\ref{stat_ansatz})
into (\ref{nls_2D_1_comp}), we obtain the stationary equation
\begin{equation}
F(\phi;\mu):=-\frac{1}{2} \nabla^{2}\phi+|\phi|^{2}\phi%
+ V(\mathbf{r})\phi-\mu\phi=0,
\label{nls_2D_1_comp_steady}
\end{equation}
where $F(\phi;\mu)$ stands for the nonlinear set of equations utilized
in our Newton solvers as well as in the deflated continuation method.
The equation is discretized with piecewise linear finite elements for
the real and imaginary components using FEniCS \cite{logg2011}.
The problem is
posed on the domain $D = (-12, 12)^2$ and homogeneous Dirichlet conditions
are imposed. This choice of domain is made to ensure that the influence
of truncating the domain is negligible, as the support of the solutions
remains far from the boundary.

We use the following diagnostic
\begin{equation}
N=\int_{\mathbb{R}^{2}} {|\phi(x,y)|^{2}}dx\,dy,
\label{noa}
\end{equation}
to summarize the parametric dependence of each steady state
branch on $\mu$. The above integral represents the number of
atoms in the BEC, considered as a function of the chemical
potential $\mu$. When $N \rightarrow 0$, the nonlinearity of
the problem becomes irrelevant and the states bifurcate from
the respective linear limit. We will also use the difference
in the number of atoms with respect to a given state $\Delta
N$ in order to highlight the emergence of bifurcations.
This diagnostic, examined in~\cite{middel2}, transparently
illustrates the origin of different branches.

In the $N\rightarrow 0$ limit, we can decompose the
modes in Cartesian form~\cite{landau_qm} as being proportional to:
\begin{eqnarray}
  \ket{m,n}_{(\textrm{c})}:=\phi_{m,n}
  \sim H_m(\sqrt{\Omega} x) H_n(\sqrt{\Omega} y) e^{-\Omega r^2/2}
\label{extra1}
\end{eqnarray}
where $H_{m,n}$ are the Hermite polynomials. 
These linear
eigenfunctions
have corresponding eigenvalues $\mu=E_{m,n}=\Omega (m + n +1)$.
On the other hand, the linear eigenfunctions can also be expressed
in polar coordinates via
\begin{eqnarray}
\ket{k,l}_{(\textrm{p})}:=\phi_{k,l} = q_{k,l}(r) e^{i l \theta}
\label{extra2}
\end{eqnarray}
with eigenvalues $\mu=E_{k,l}= (1 + |l| + 2 k) \Omega$.
The parameters $l$ and $k$ denote the eigenvalue of the ($z$-component of the) angular
momentum operator and the 
number of radial zeros of the corresponding eigenfunction respectively.
This radial part of the relevant eigenfunction is given by~\cite{brand}
\begin{eqnarray}
q_{k,l} \sim r^l L_k^l(\Omega r^2) e^{-\Omega r^2/2},
\label{extra3}
\end{eqnarray}
where $L_k^l$ are the associated Laguerre polynomials. The subscripts in $\ket{m,n}_{(\textrm{c})}$
and $\ket{k,l}_{(\textrm{p})}$ stand for the Cartesian (c) 
and polar (p) representations respectively. 
In most of what follows, we 
will use the Cartesian notation, but we will occasionally
resort to the polar decomposition where convenient.

Once the relevant solutions have been
identified, we wish to investigate their stability. We
assume the perturbation ansatz around a stationary solution
$\phi^{0}$ to be of the following form:
\begin{equation}
\widetilde{\Phi }(x,y,t)=e^{-i\mu t}\Big\lbrace \phi^{0}(x,y)+\varepsilon
\left[a(x,y)e^{i\omega t}+b^{\ast}(x,y)e^{-i\omega^{\ast} t}\right]\Big\rbrace,\
\label{lin_ansatz}
\end{equation}
where $\omega$ is the (complex) eigenfrequency, $\varepsilon $ is a 
(formal) small 
amplitude of the perturbation, and the asterisk stands for complex
conjugation. Inserting Eq.~(\ref{lin_ansatz}) into the NLS equation
(\ref{nls_2D_1_comp}), we obtain at order $\varepsilon$ an eigenvalue 
problem written in the following matrix form:
\begin{equation}
\rho
\begin{pmatrix}
a \\
b \\
\end{pmatrix}
=
\begin{pmatrix}
A_{11} & A_{12} \\
-A_{12}^{\ast } & -A_{11} 
\end{pmatrix}%
\begin{pmatrix}
a\\
b\\
\end{pmatrix}
,  \label{eig_prob}
\end{equation}
with eigenfrequencies $\rho=-\omega$ (the eigenvalues $\lambda=i \omega$), 
eigenvectors $\mathcal{V}=(a,b)^{T}$, and matrix elements given by
\begin{subequations}
\begin{eqnarray}
A_{11} &=&-\frac{1}{2}\nabla^{2}+2|\phi^{0}|^{2}+V(x,y)-\mu, \\
A_{12} &=&\left(\phi^{0}\right)^{2}. 
\end{eqnarray}
\end{subequations}
The steady state $\phi^{0}$ is classified as stable in this
Hamiltonian system if no
eigenfrequency $\omega =\omega_{r}+i\,\omega _{i}$ has a non-vanishing
imaginary part $\omega _{i}$; this is because, given the Hamiltonian
nature, if $\omega$ is an eigenfrequency, so is $-\omega$, $\omega^{\star}$
and $-\omega^{\star}$. The scenario of stability will be
depicted by a solid blue 
line in the bifurcation diagrams presented below. On the other hand, 
when the solution becomes unstable, two
types of instabilities can be identified: i) \textit{exponential instabilities}
characterized by a pair of imaginary eigenfrequencies with \textit{zero} 
real part, and ii) \textit{oscillatory instabilities} characterized by a 
complex eigenfrequency quartet. These two scenarios are depicted
by dashed-dotted red and green lines respectively in the bifurcation diagrams
that follow to highlight the nature of the dominant 
unstable mode; a transition between these two colors will thus
signal a change in the dominant instability type.

We are now ready to describe the different solutions produced by the deflation technique.

\section{Numerical Results}

The simplest state of the system is its ground
state $|0,0 \rangle_{(\textrm{c})}$ with eigenvalue $\mu=\Omega$
at the linear limit. As this state is generically stable~\cite{book_new}
(i.e., for all values of $\mu$ from the linear limit to the
Thomas-Fermi regime), no bifurcations
occur from it. This solution is well-known and we do not
examine it further.

\subsection{Bifurcations from $\mu = 2\Omega$}
\begin{figure}[tbp]
\begin{center}
\includegraphics[height=.16\textheight, angle =0]{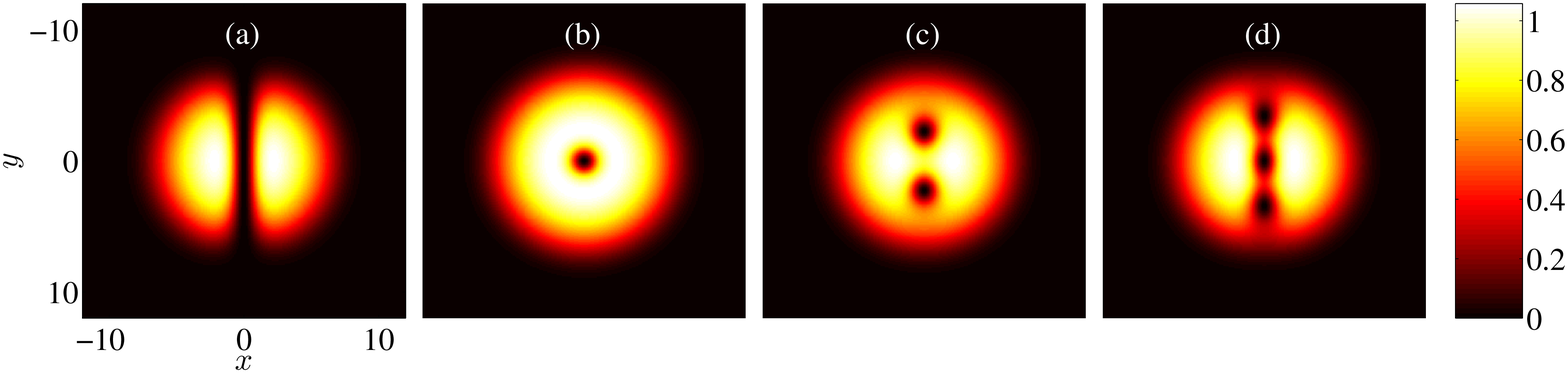}
\includegraphics[height=.16\textheight, angle =0]{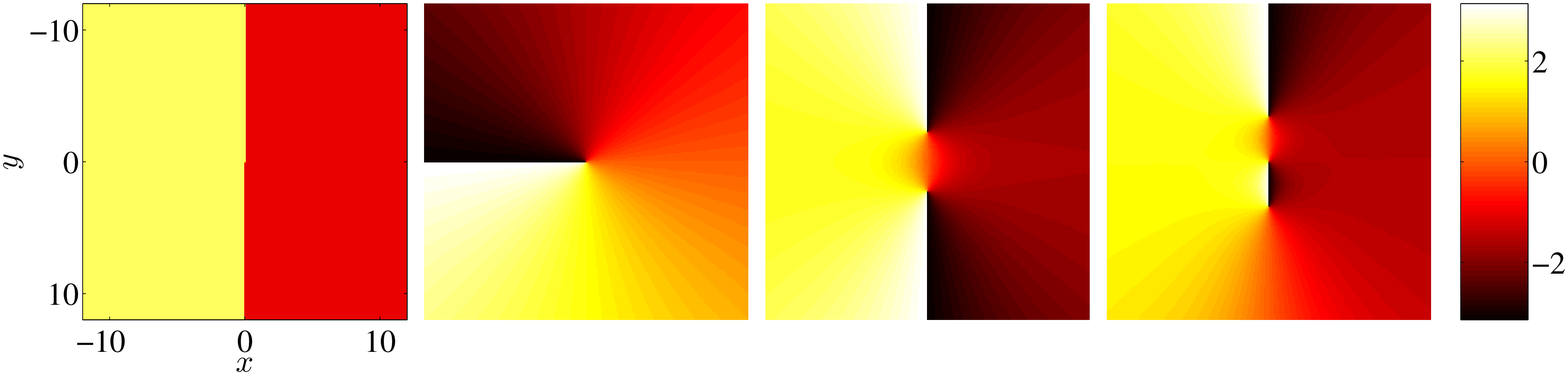}
\includegraphics[height=.16\textheight, angle =0]{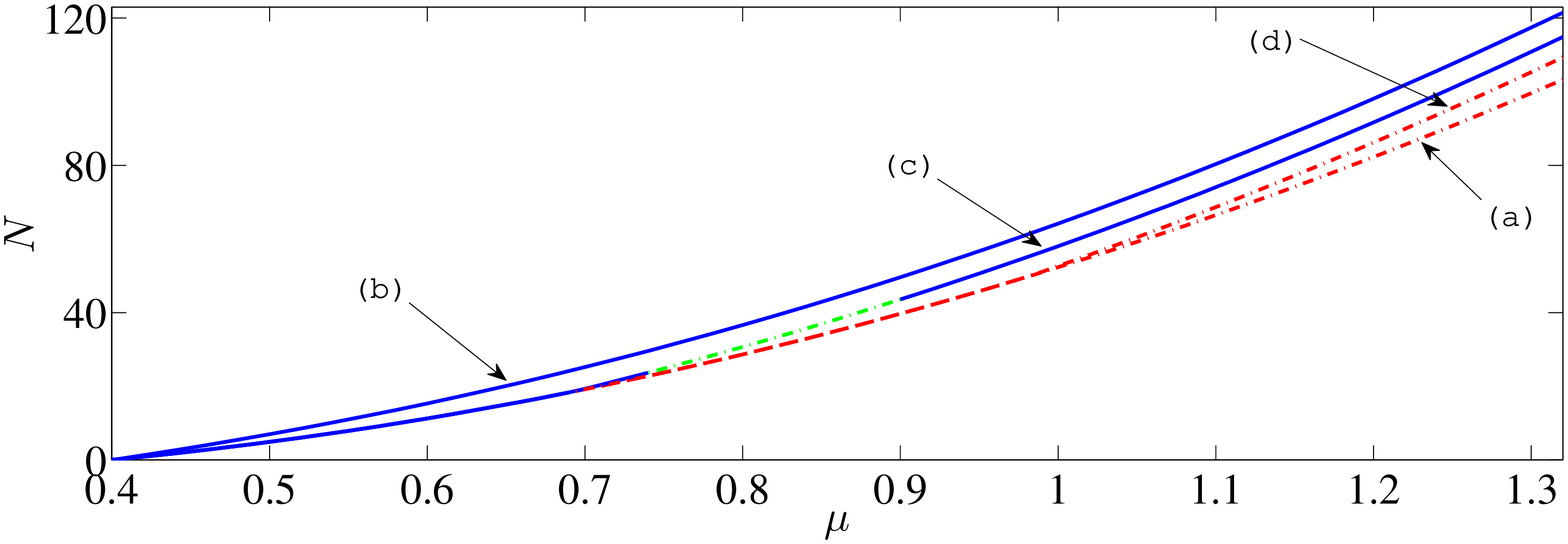}
\includegraphics[height=.16\textheight, angle =0]{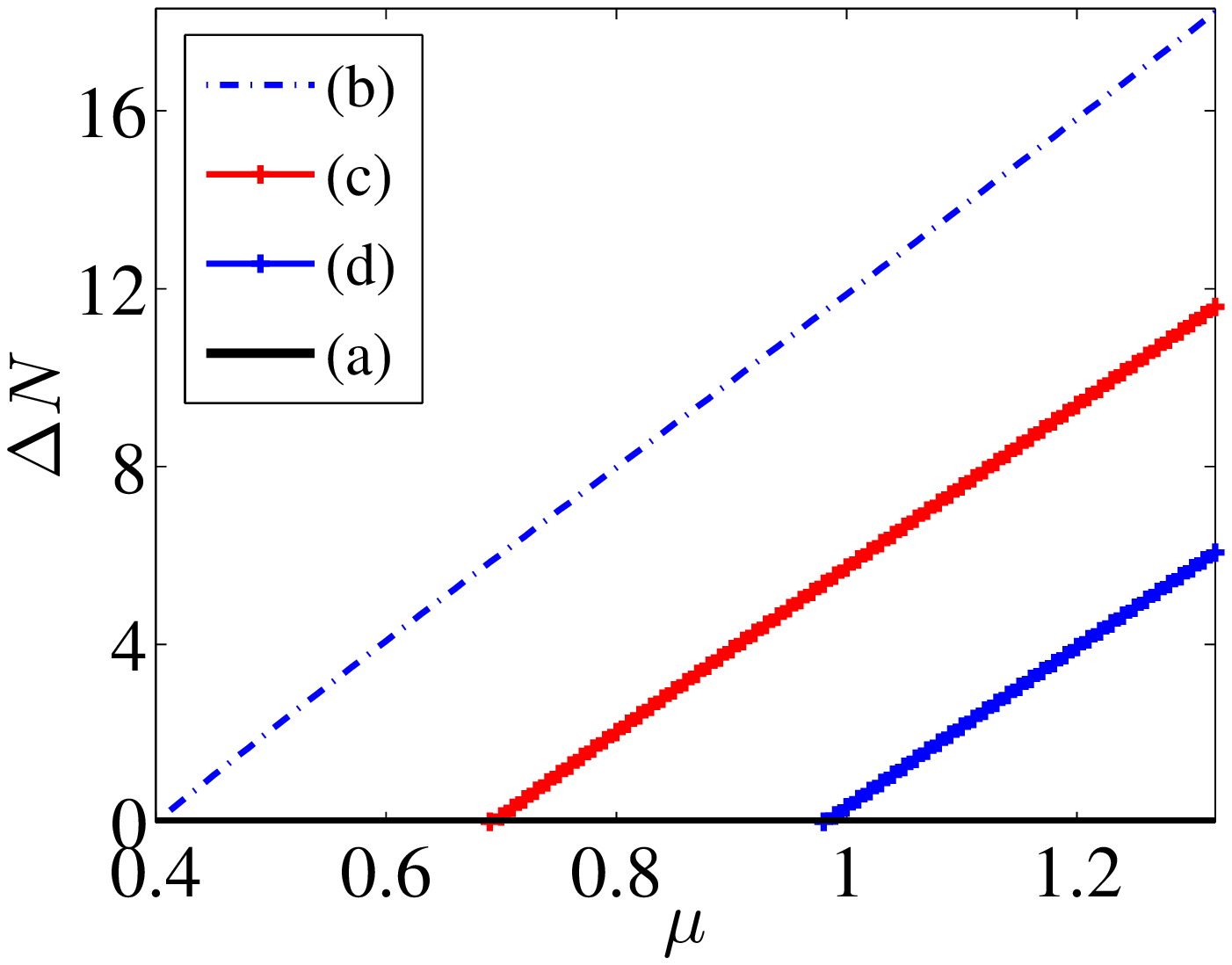}
\end{center}
\caption{
(Color online) A bifurcation diagram of states emanating
from the trivial branch at $\mu = 2\Omega$.  Top and middle
rows correspond to plots of the density profiles and
phases, respectively, of the (a) dark soliton stripe ($\mu=1.2$),
(b) single vortex ($\mu=1.32$), (c) vortex dipole ($\mu=1.23$),
and (d) vortex tripole ($\mu=1.23$) states. Bottom panels correspond
to the number of atoms $N$ (left) and atom number difference 
$\Delta N$ (right) from the dark soliton stripe branch , as 
functions of $\mu$. The secondary bifurcations occur at 
$\mu\approx 0.68$ and $\mu\approx0.98$.
}
\label{fig2}
\end{figure}

From the point of view of bifurcation analysis, the first
interesting events occur at $\mu = 2\Omega$, with $n + m =
1$. As is well-known~\cite{middel1,middel2,azpeitia}, two branches bifurcate
from this point (Fig.~\ref{fig2}).  One is the dark
soliton stripe $|1,0 \rangle_{(\textrm{c})}$,
Fig.~\ref{fig2}(a).  The other is the single charge (i.e., unit vorticity)
vortex of Fig.~\ref{fig2}(b). In Cartesian coordinates this
is described as the linear combination $|1,0
\rangle_{(\textrm{c})} + i |0,1 \rangle_{(\textrm{c})}$; in
polar coordinates it is $|0,1 \rangle_{(\textrm{p})}$.
Among the two, the vortex is very robust, incurring no
instabilities \cite{middel1,book_new}.  On the other hand,
as the chemical potential increases the stripe progressively
approaches a rectilinear dark soliton that is well-known to
be subject to a transverse (modulational)
instability~\cite{kuzne}.  In fact, there is a whole cascade
of such instabilities, arising in the form of pitchfork
bifurcations from the stripe \cite{middel1,middel2,book_new}.  The first
of these bifurcations gives rise to the emergence of the
vortex dipole state, Fig.~\ref{fig2}(c). This state is
well-known and has been studied
experimentally~\cite{bpanderson,ushall}.  It is
dynamically stable except for a narrow interval of
oscillatory instability (associated with a Hamiltonian Hopf
bifurcation), as previously noted~\cite{middel1,middel2}.
The next bifurcation gives rise to a vortex tripole,
Fig.~\ref{fig2}(d). By this stage the stripe branch is
unstable and the vortex tripole inherits this instability.
This is a configuration with three vortices of alternating
charge $(+,-,+)$ or $(-,+,-)$. This has also been identified
in experiments~\cite{bagnato} and explored in
simulations~\cite{vas1,vas2}. This pattern of bifurcations
continues to higher excited states for larger values of
$\mu$, yet we do not pursue these bifurcations further,
given their earlier analysis, e.g., in~\cite{middel1,middel2}.

\subsection{Bifurcations from $\mu = 3\Omega$}

\begin{figure}[tbp]
\begin{center}
\includegraphics[height=.16\textheight, angle =0]{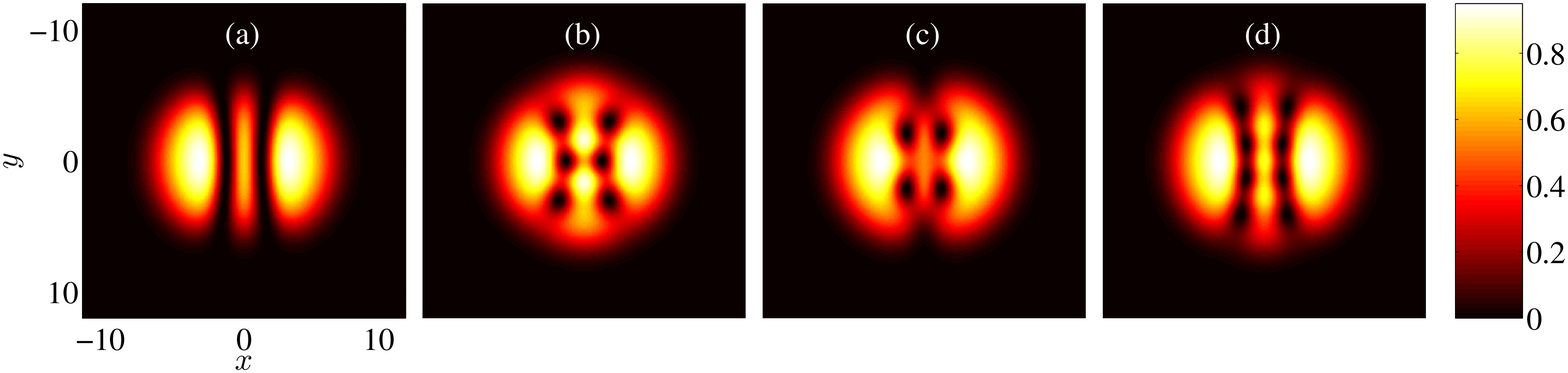}
\includegraphics[height=.16\textheight, angle =0]{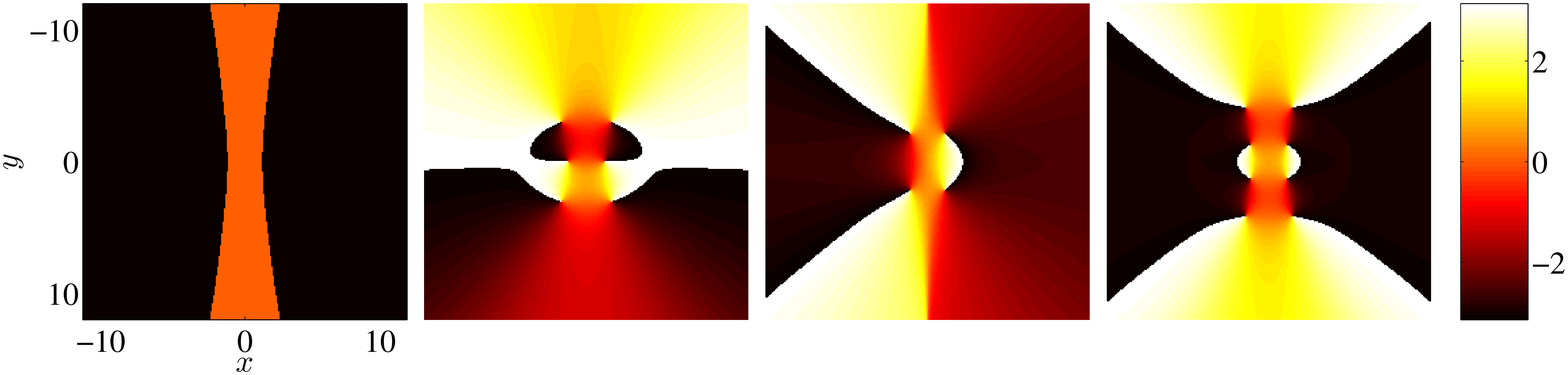}
\includegraphics[height=.16\textheight, angle =0]{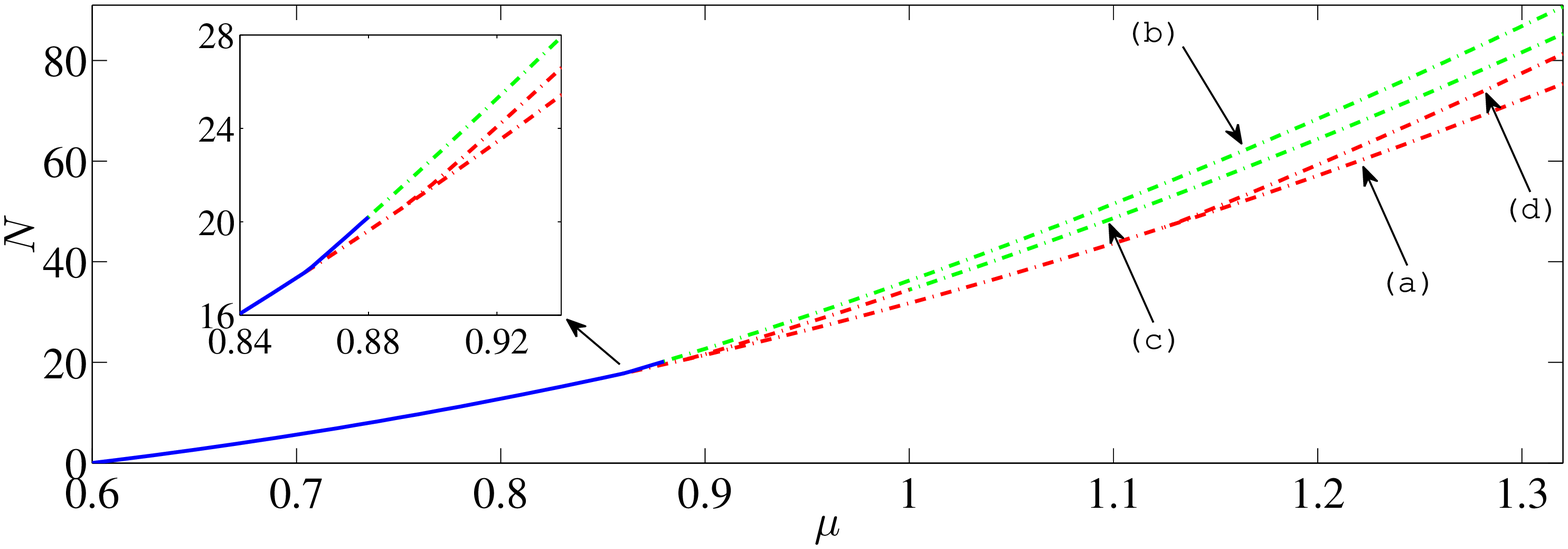}
\includegraphics[height=.16\textheight, angle =0]{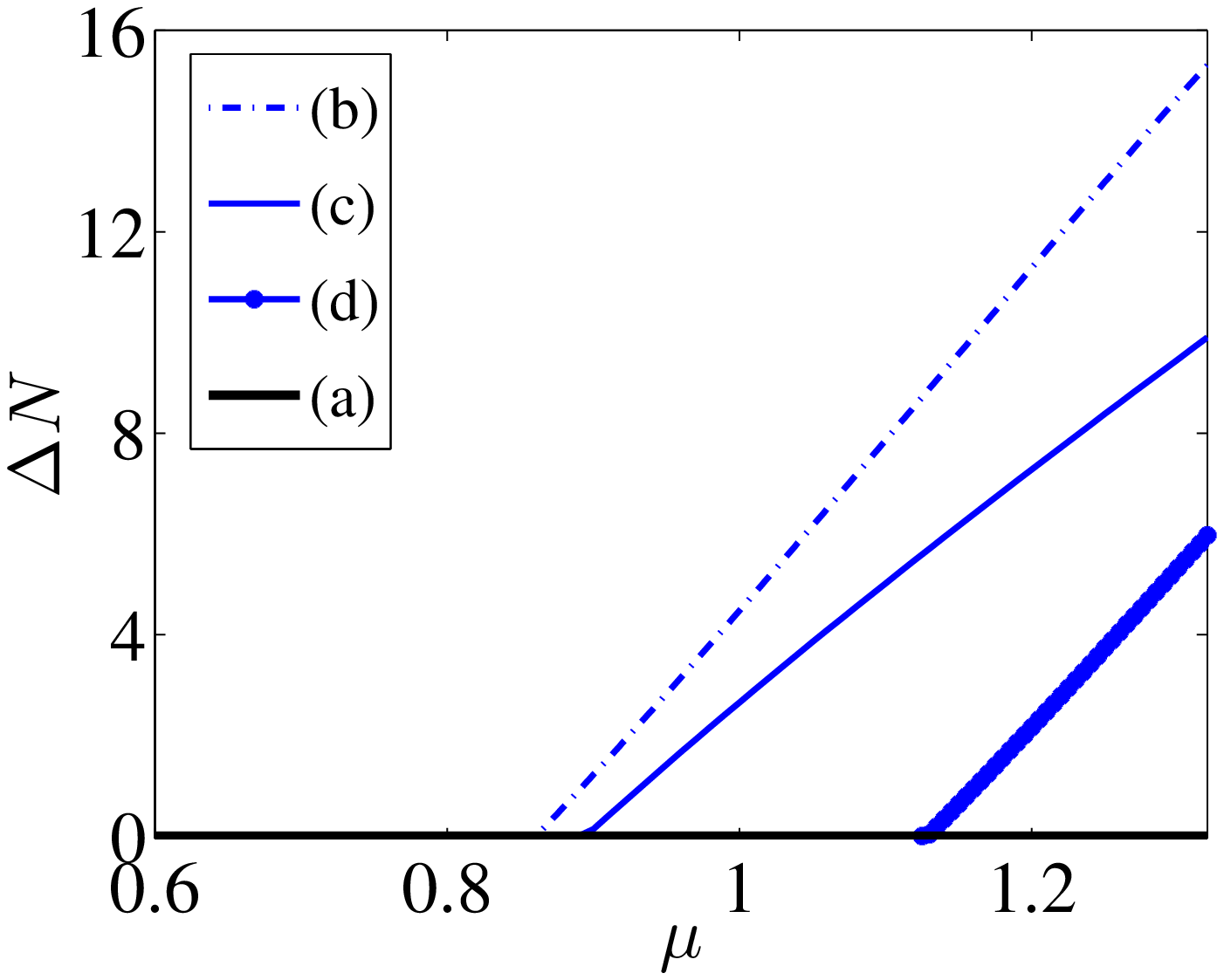}
\end{center}
\caption{
(Color online) Bifurcations from the second excited state
$|2,0 \rangle_{(\textrm{c})}$. Top and middle panels correspond
to plots of the density profiles and phases, respectively, of the 
(a) two dark soliton stripes ($\mu=1.2$), (b) aligned six vortex 
($\mu=1.15$), (c) double vortex dipole ($\mu=1.2$), and (d) aligned
eight vortex ($\mu=1.27$), state. Bottom panels correspond to the number
of atoms $N$ (left) and atom number difference $\Delta N$ (right) from 
the two dark soliton stripes branch, as functions of $\mu$. State (a) 
exists from the linear limit; branches (b), (c) and (d) emerge at values
of $\mu$ of $\mu\approx0.86$, $\mu\approx0.89$ and $\mu\approx1.12$, respectively.
}
\label{fig4}
\end{figure}

\begin{figure}[tbp]
\begin{center}
\includegraphics[height=.158\textheight, angle =0]{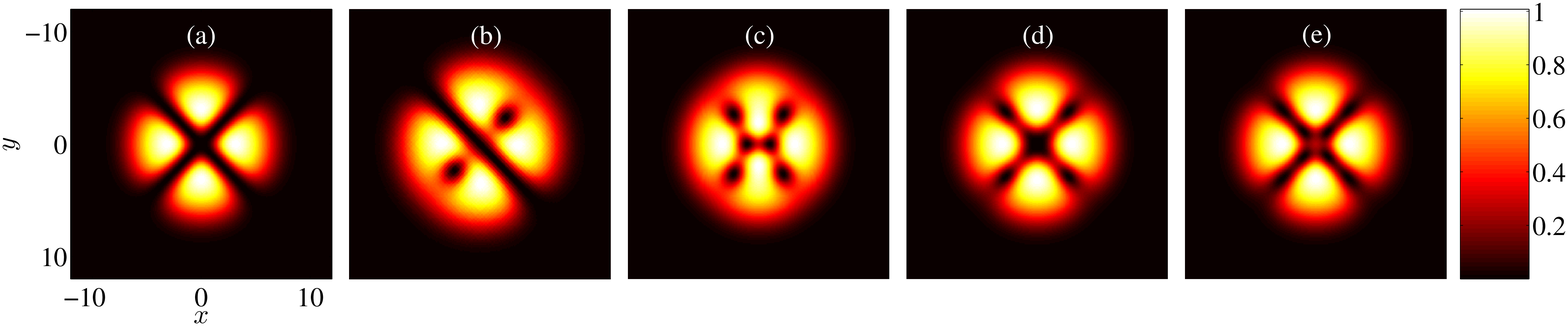}
\includegraphics[height=.158\textheight, angle =0]{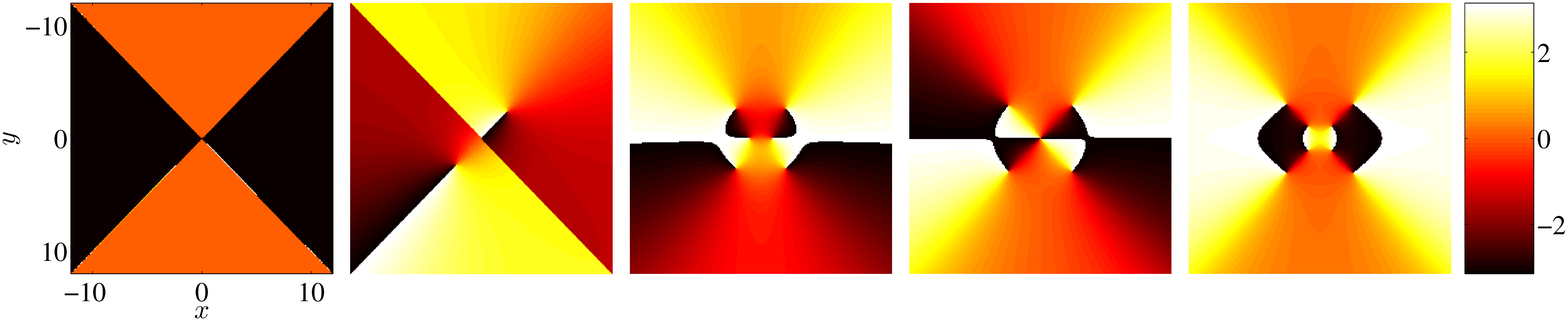}
\includegraphics[height=.16\textheight, angle =0]{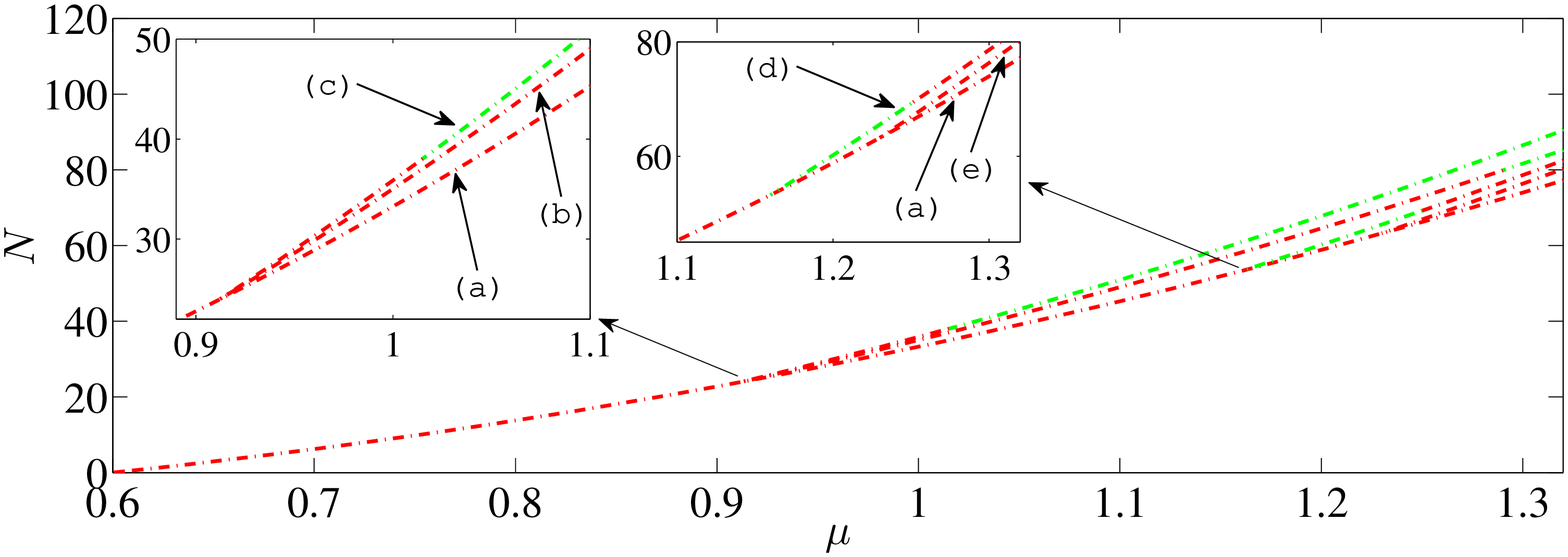}\includegraphics[height=.153\textheight, angle =0]{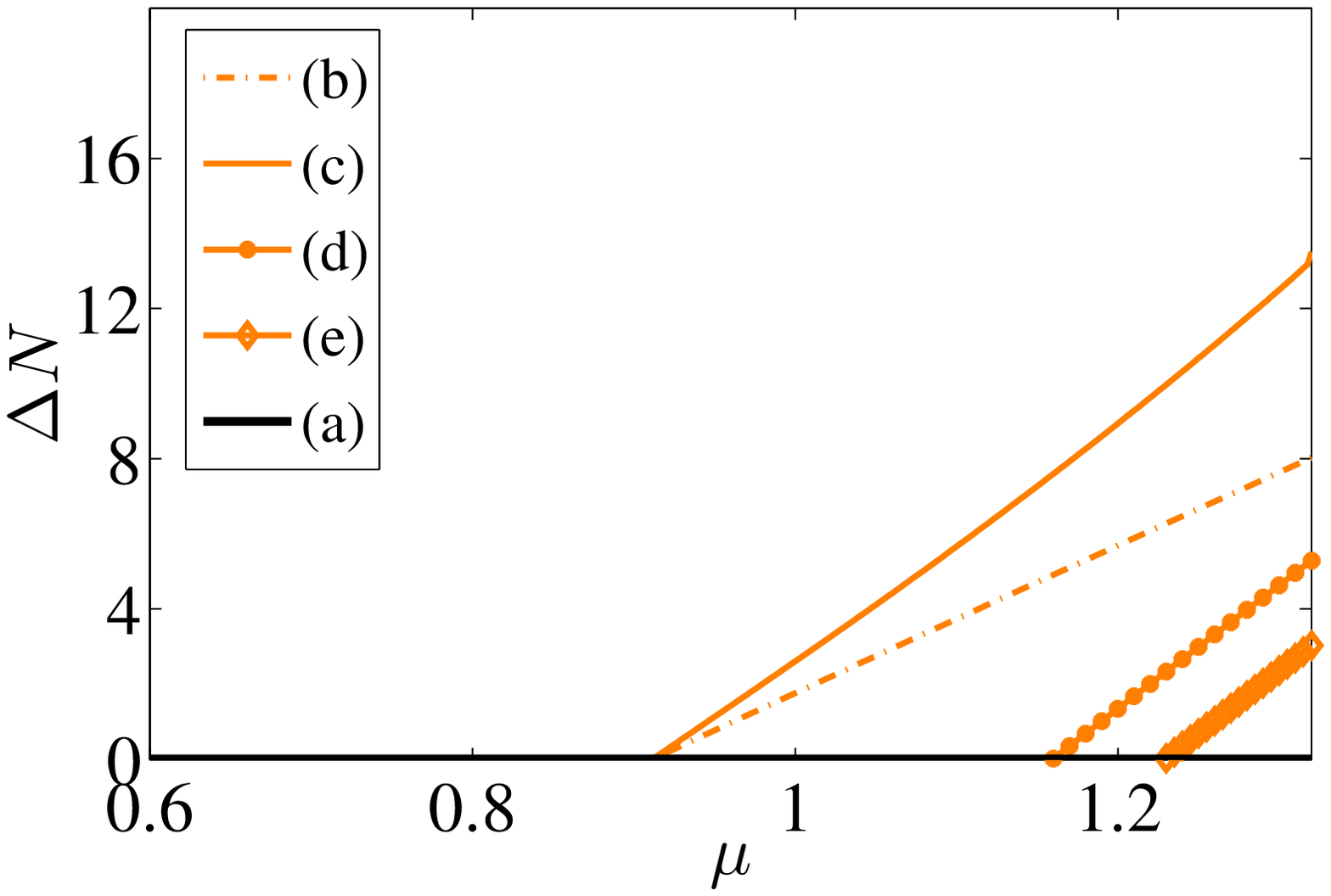}
\end{center}
\caption{
(Color online) Same as the previous figure, but
for the $|1,1 \rangle_{(\textrm{c})}$ branch. Top and middle panels 
correspond to plots of the density profiles and phases, respectively, 
of the (a) dark soliton cross i.e., the $|1,1 \rangle_{(\textrm{c})}$ 
state ($\mu=1.23$), (b) dark soliton stripe together with two same 
charge vortices ($\mu=1.23$), (c) diagonal six vortex ($\mu=1.3$),
(d) diagonal five vortex state, consisting of one vortex of charge $l=-2$ surrounded by four vortices of $l=1$ 
($\mu=1.32$), and (e) diagonal eight vortex ($\mu=1.32$), state. The bottom 
panels correspond to the number of atoms $N$ (left) and atom number 
difference $\Delta N$ (right) from the two dark soliton cross branch,
as functions of $\mu$. State (a) exists from the linear limit; branches
(b), (c), (d) and (e) emerge at values of $\mu$ of $\mu\approx0.91$, 
$\mu\approx0.912$, $\mu\approx1.16$, and $\mu\approx1.23$, respectively.
}
\label{fig5}
\end{figure}

\begin{figure}[tbp]
\begin{center}
\includegraphics[height=.158\textheight, angle =0]{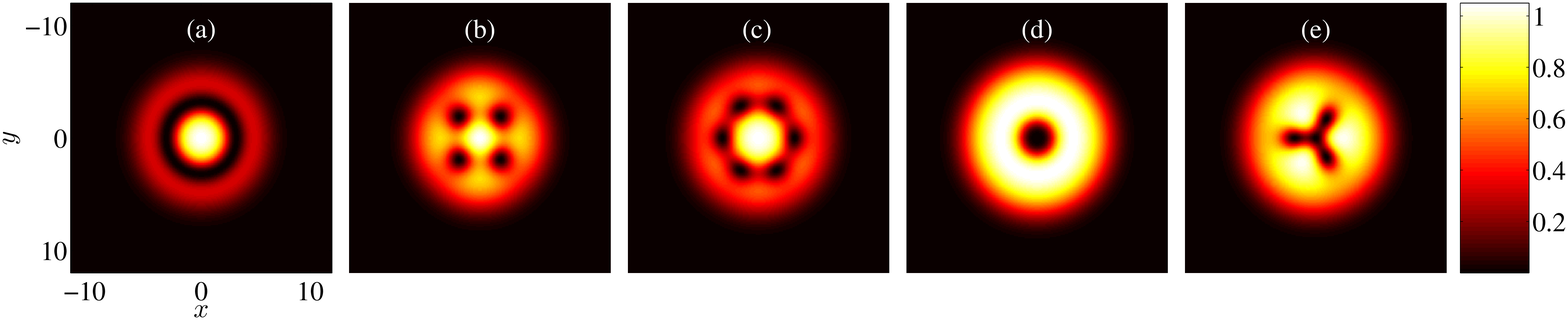}
\includegraphics[height=.158\textheight, angle =0]{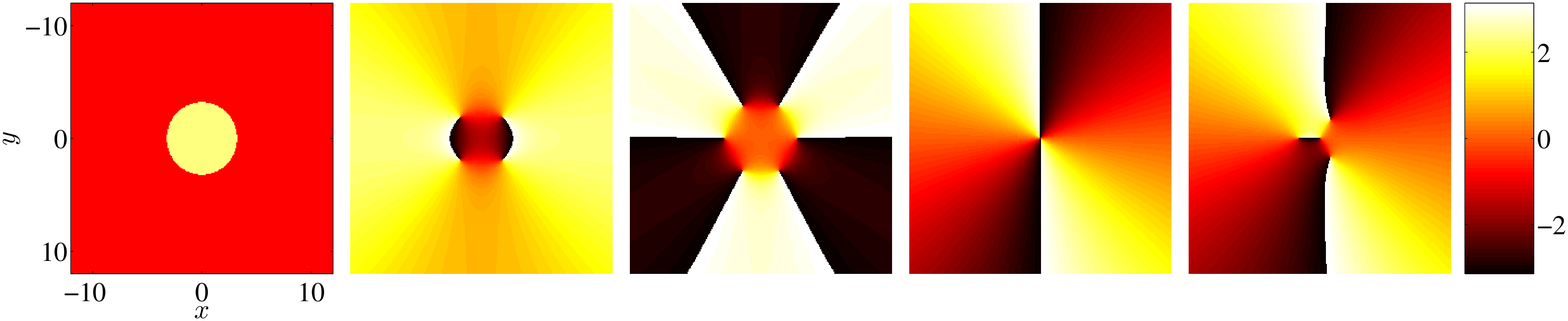}
\includegraphics[height=.16\textheight, angle =0]{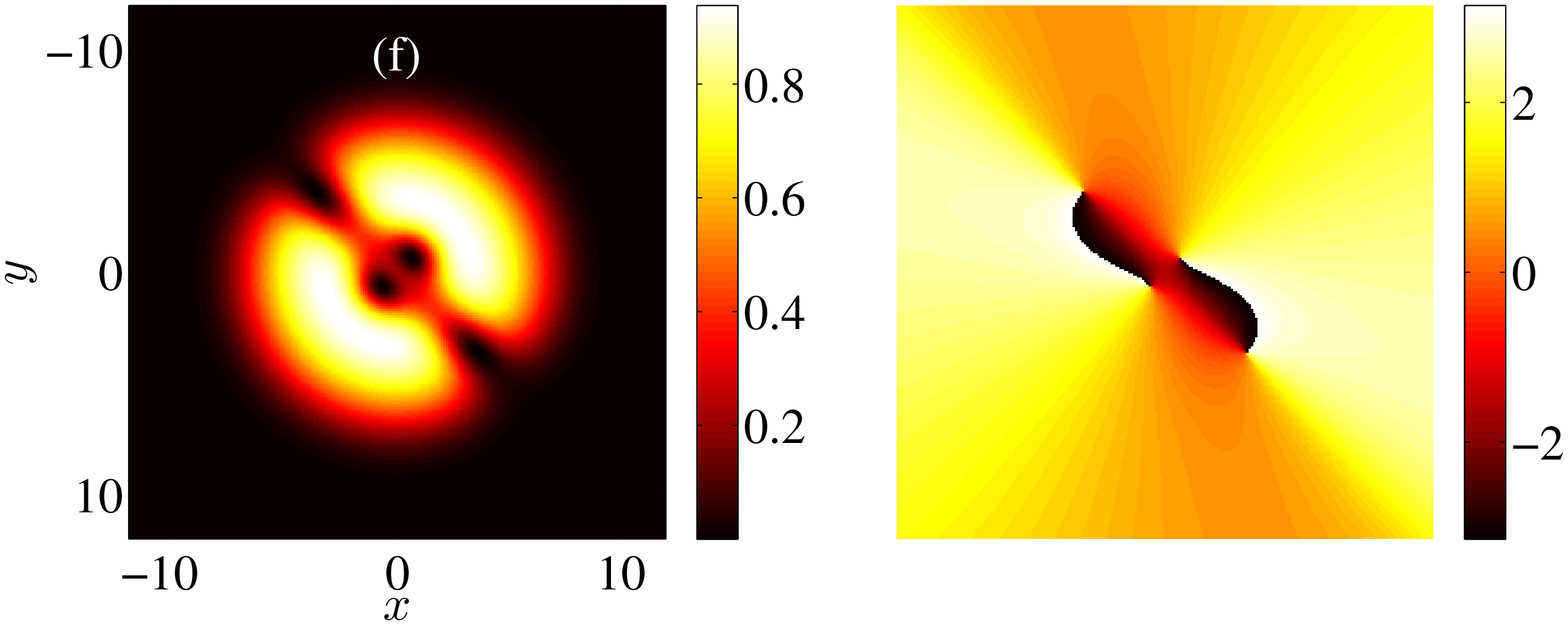}\includegraphics[height=.16\textheight, angle =0]{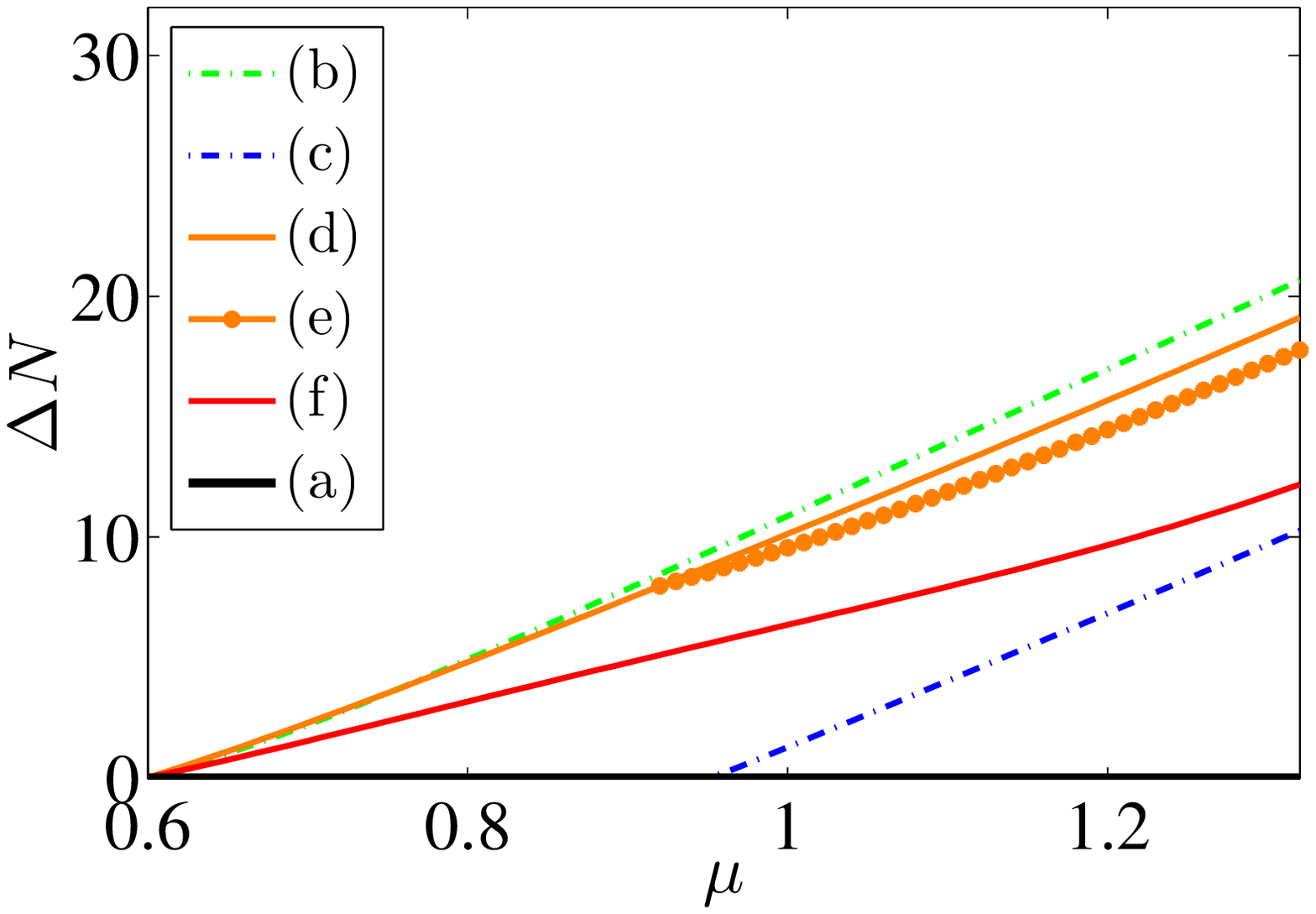}
\includegraphics[height=.18\textheight, angle =0]{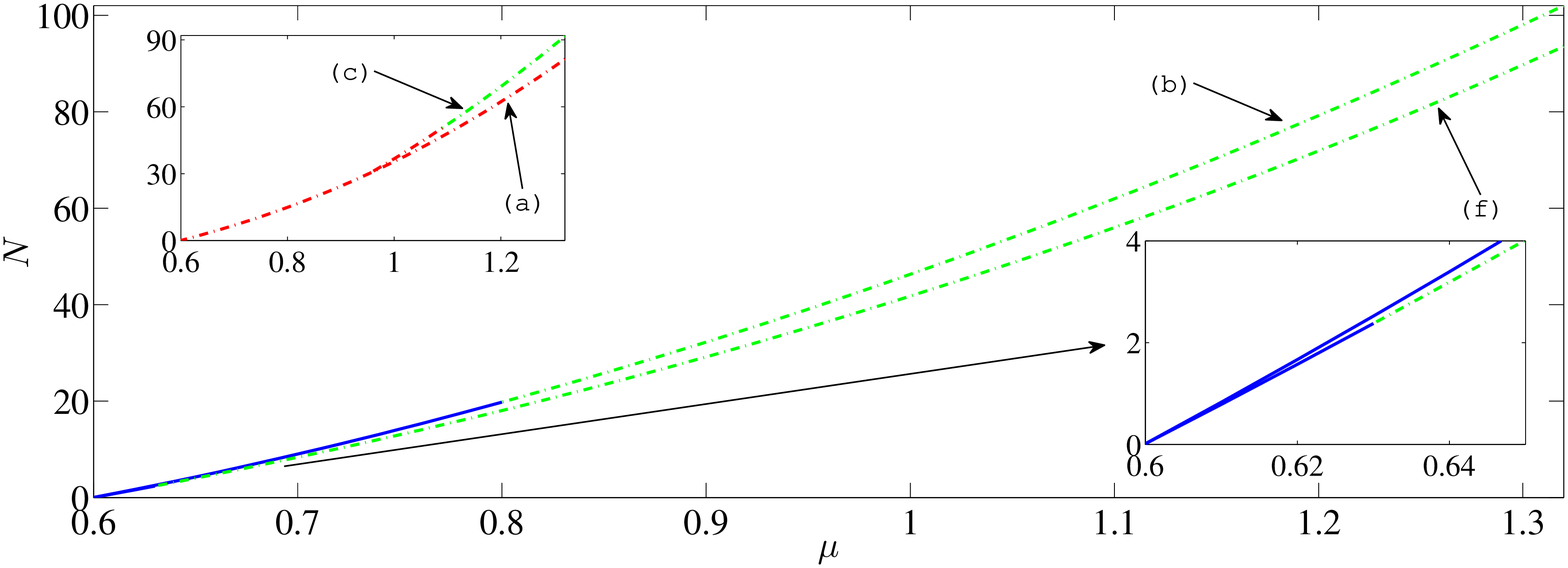}
\includegraphics[height=.18\textheight, angle =0]{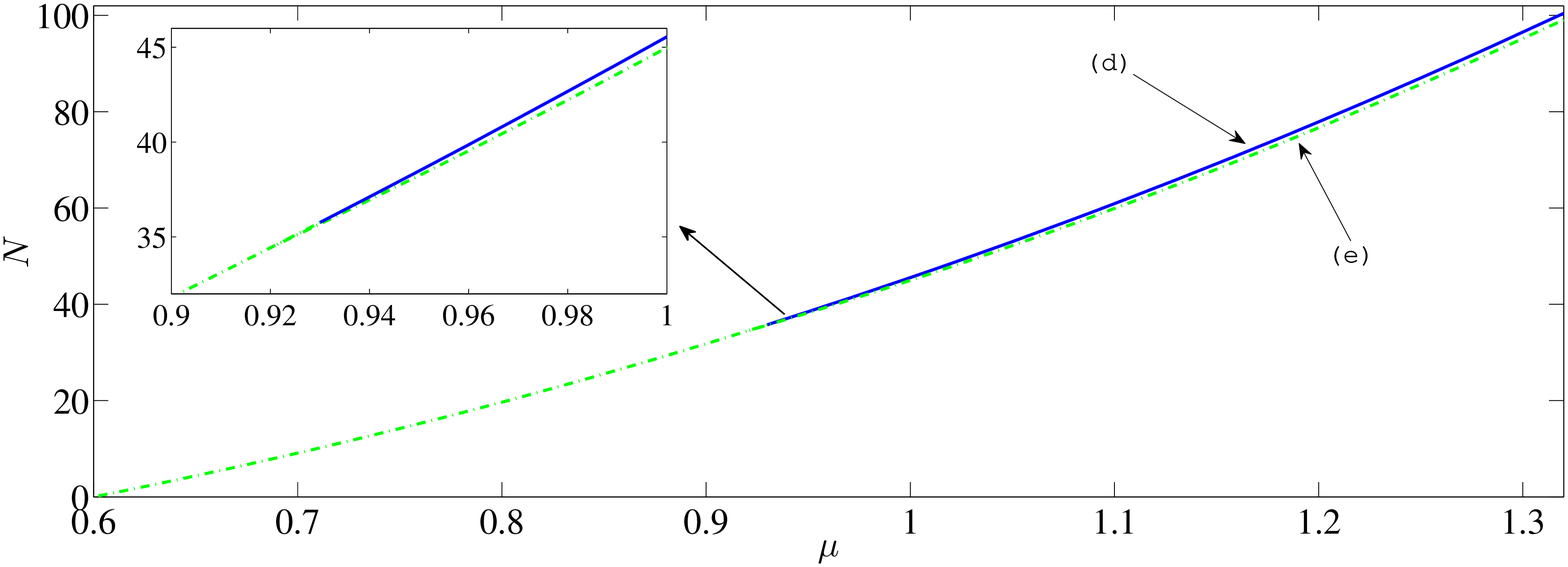}
\end{center}
\caption{
(Color online) Same as Fig.~\ref{fig2} but for states bifurcating
from $\mu=3\Omega$. First and second rows as well as the left panels
in the third row correspond to plots of the density profiles and phases,
respectively, of the (a) ring dark soliton ($\mu=1.0$), (b) vortex quadrupole,
which bifurcates from the linear limit and is dynamically stable except for an
oscillatory instability interval ($\mu=1.1$), (c)
vortex hexagon (for $\mu=1.23$), which bifurcates through a symmetry breaking
instability of a ring dark soliton (a), (d) charge-two vortex ($\mu=1.23$),
subject to a sequence of oscillatory instabilities, (e) three same charge
vortices surrounding an oppositely charged one ($\mu=1.32$),
(f) $\ket{2,0}_{(\textrm{c})}+i\ket{1,1}_{(\textrm{c})}$ ($\mu=1.32$),
state. The right panel in the third row as well as the panels in the 
fourth and fifth rows correspond to the atom number difference $\Delta N$
from the two ring dark soliton branch, and the number of atoms $N$, respectively,
all as functions of $\mu$. Note that bifurcations happen for $\mu\approx 0.92$
and $\mu\approx0.95$, respectively, for the branches (e) and (c).
}
\label{fig3}
\end{figure}

We now turn to the considerably more complicated case of
bifurcations from $\mu = 3\Omega$ with $n + m = 2$. These
bifurcations are also mostly well-known (although with 
some important twists to be discussed below), and are summarized
in Figs.~\ref{fig4}-\ref{fig3}. Fig.~\ref{fig4} considers
$|2,0 \rangle_{(\textrm{c})}$ and its subsequent bifurcations. 
Just as a single soliton stripe gave rise to a single vortex dipole 
in Fig.~\ref{fig2}, a double soliton stripe gives rise to a double 
vortex tripole in Fig.~\ref{fig4}(b), a double vortex dipole
in  Fig.~\ref{fig4}(c) and a double aligned vortex quadrupole
in Fig.~\ref{fig4}(d). Somewhat surprisingly, the double tripole
bifurcation (leading to a state with 6 vortices of alternating
charges) corresponds to a bifurcating stable branch,
at least when the latter states first emerge; for
higher $\mu$ they become oscillatorily unstable, acquiring
a complex eigenfrequency quartet. The subsequent bifurcations
arise from the already unstable double soliton stripe branch
and hence the bifurcating branches are also unstable. For a
demonstration of the relevant stability properties, see the
bottom left panel of Fig.~\ref{fig4}, while the bottom
right uses the number of atom difference from the two-soliton
branch as a diagnostic to display the occurrences of the different bifurcations
(as the daughter branches depart from zero in this
quantity).

The bifurcation of the 6 vortex state (numerically
occurring at $\mu=0.86$;
panel ~\ref{fig4}(b))
precedes that of the 4 vortex one (numerically
occurring around $\mu=0.89$;
panel ~\ref{fig4}(c)). The former bifurcation in the  formulation
of~\cite{middel2} comes from the combination of  $|2,0\rangle_{(\textrm{c})}$
with  $|0,3\rangle_{(\textrm{c})}$, with a $\pi/2$ phase shift,
while the latter emerges from the symmetry-breaking event involving
$|2,0\rangle_{(\textrm{c})}$
with  $|1,2\rangle_{(\textrm{c})}$ (again with a $\pi/2$ phase shift).
It is also intriguing that
the theoretical prediction of these bifurcations based on the
two mode theory of~\cite{middel2} occurs at $503 \Omega/113 \approx 0.89$
and $265 \Omega/61 \approx 0.87$, respectively, i.e., very close
to the computationally obtained values, although in reverse order.
%

Fig.~\ref{fig5} considers $|1,1\rangle_{(\textrm{c})}$ and its subsequent 
bifurcations. The first bifurcation from $|1,1\rangle_{(\textrm{c})}$
gives rise to a branch which features a dark soliton stripe together with 
two same charge vortices, as shown in
Fig.~\ref{fig5}(b). Nearly concurrent to this bifurcation is the
emergence of a branch involving a ``diagonal'' set of 6 vortices;
see Fig.~\ref{fig5}(c).
As discussed in~\cite{middel2}, the branch of solutions in
Fig.~\ref{fig4}(b) collides and disappears (in a saddle-center
bifurcation) with that of Fig.~\ref{fig5}(c), as $\mu$ is increased.
A subsequent bifurcation gives
rise to a branch with a vortex of charge $l=-2$ in the middle and four 
surrounding vortices of charge $l=1$, Fig.~\ref{fig5}(d). Explicit algebraic
conditions for such states have been obtained via a generating function approach
in the large-density Thomas--Fermi limit \cite{generating}, as have the 
rectilinear vortex states of Fig.~\ref{fig2}. It is also interesting to note,
as illustrated in the bottom panel of Fig.~\ref{fig5}, the resulting branches
out of these bifurcations also feature intervals of oscillatory
instabilities. However, the principal branch $|1,1\rangle_{(\textrm{c})}$
out of which these bifurcations arise is unstable,
and hence all the bifurcating branches inherit this instability.
This includes the 8 diagonal vortex state~Fig.~\ref{fig5}(e).

A more complex structure emerges in the context of
solutions with radially symmetric density and their
bifurcations that are analyzed in Fig.~\ref{fig3}. 
The first such solution to consider is the ring dark
solution state, Fig.~\ref{fig3}(a), which is well-known and
has been extensively studied (for a recent discussion,
see~\cite{wenlong} and the references therein). In Cartesian
notation, this is $|2,0 \rangle_{(\textrm{c})} + |0,2
\rangle_{(\textrm{c})}$, while in radial notation this is
the $\ket{1,0}_{(\textrm{p})}$ state.  A systematic study of
the stability of this and related states from the linear
limit onwards was conducted in~\cite{toddrotating},
showing that its degeneracy with the vortex quadrupole of
Fig.~\ref{fig3}(b) leads to an immediate quadrupolar
instability through a real eigenvalue pair for this mode. In
contrast, the vortex quadrupole is generically
stable, aside from a finite interval of oscillatory
instability~\cite{middel2}. The ring dark soliton,
progressively becomes more unstable to undulations with
higher wavenumbers. The first to emerge is a hexapolar mode,
giving rise (through a pitchfork bifurcation) to the vortex
hexagons of Fig.~\ref{fig3}(c); these may for larger values
of $\mu$ also possess oscillatory instabilities (as denoted
in the bottom panel). This pattern continues with
an octapolar mode leading to vortex octagons and so on. 

Another mode bifurcating here from the linear limit is the doubly
charged vortex $\ket{0,2}_{(\textrm{p})}$ of Fig.~\ref{fig3}(d). 
This state is unstable from the linear limit onwards through a 
sequence of intervals of oscillatory instabilities originally 
examined in~\cite{pu99} and subsequently retraced in a variety
of publications~\cite{herring,pegokollar}.
This state can also be represented via a combination of Cartesian
eigenstates as $|2,0 \rangle_{(\textrm{c})} - |0,2\rangle_{(\textrm{c})}+ 2 i |1,1 \rangle_{(\textrm{c})}$.
Out of this branch bifurcates branch (e) which bears three vortices
of the same charge in the periphery and one of opposite charge at
the center (hence has the same total charge of $2$). Intriguingly, 
due to the spherical symmetry of the solution, despite the pitchfork
nature of the bifurcation, the branch (e) has a pair of eigenvalues
of the linearization at the origin, and no genuinely imaginary
eigenfrequencies.
It is also interesting to mention here that the bifurcation of (e)
essentially coincides (for our parametric resolution of
steps of $0.01$ in the chemical potential) with the stabilization
against oscillatory instabilities of the charge $2$ branch.
Finally, the branch (f) emanating from the linear
limit and being subject to oscillatory instabilities can be
approximated by $\ket{2,0}_{(\textrm{c})}+i\ket{1,1}_{(\textrm{c})}$. 
This solution also seems to ``harbor'' 4 vortices, although 2 are 
more clearly observable close to the condensate center, while the 
two others are less discernible, merging with the background.
A particularly intriguing feature of the latter branch of solutions 
is that the systematic classification of~\cite{toddrotating} identified
the solutions stemming from the linear limit (including the ring of 
Fig.~\ref{fig3}(a), the multipole of Fig.~\ref{fig2}(b), the soliton 
necklace of Fig.~\ref{fig5}(a), the radially symmetric vortex of 
Fig.~\ref{fig3}(d) and the vortex necklace of Fig.~\ref{fig3}(b)). 
A remarkable feature of our analysis is 
that the branch of Fig.~\ref{fig3}(f) appears to have never been previously
discussed, to the best of our knowledge. This branch is subject to
oscillatory instabilities, as shown in the bottom panel of Fig.~\ref{fig3}.

The analysis presented heretofore considered bifurcations
that have been analyzed in earlier works; we have successfully
validated the method by comparing it to the known bifurcation 
diagram of this problem and in the process have unraveled novel
branches of solutions, such as Fig.~\ref{fig4}(c),~\ref{fig5}(b)
and Fig.~\ref{fig3}(f).


\subsection{Bifurcations from $\mu = 4\Omega$}

\begin{figure}[tbp]
\begin{center}
\includegraphics[height=.158\textheight, angle =0]{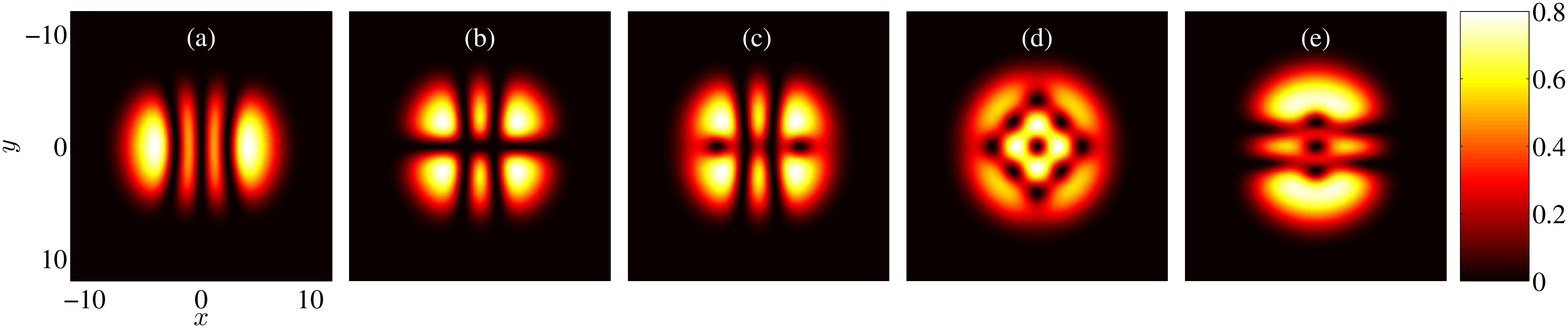}   
\includegraphics[height=.158\textheight, angle =0]{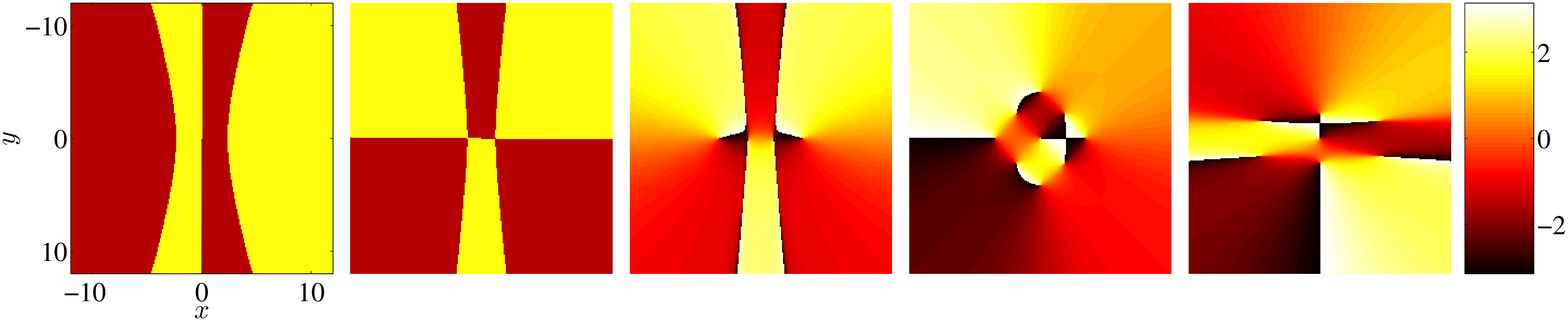}
\includegraphics[height=.158\textheight, angle =0]{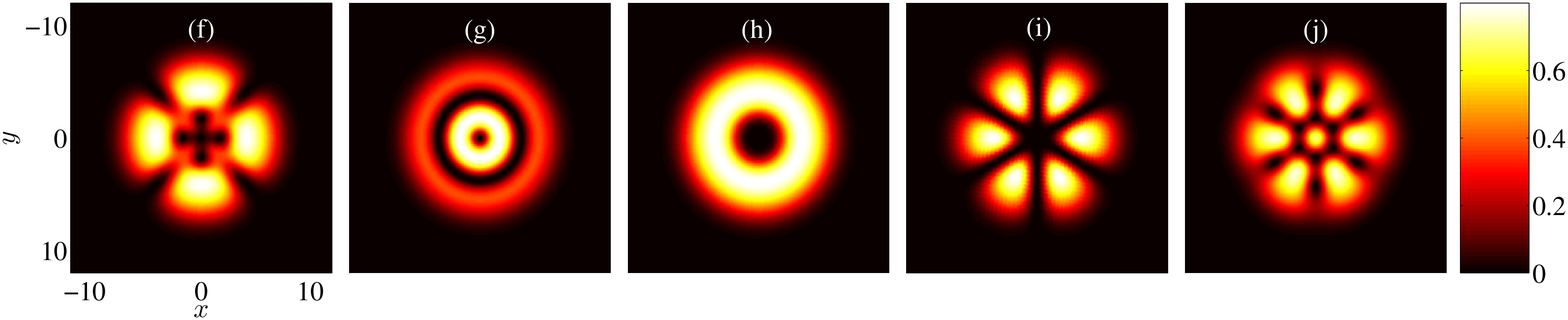}   
\includegraphics[height=.158\textheight, angle =0]{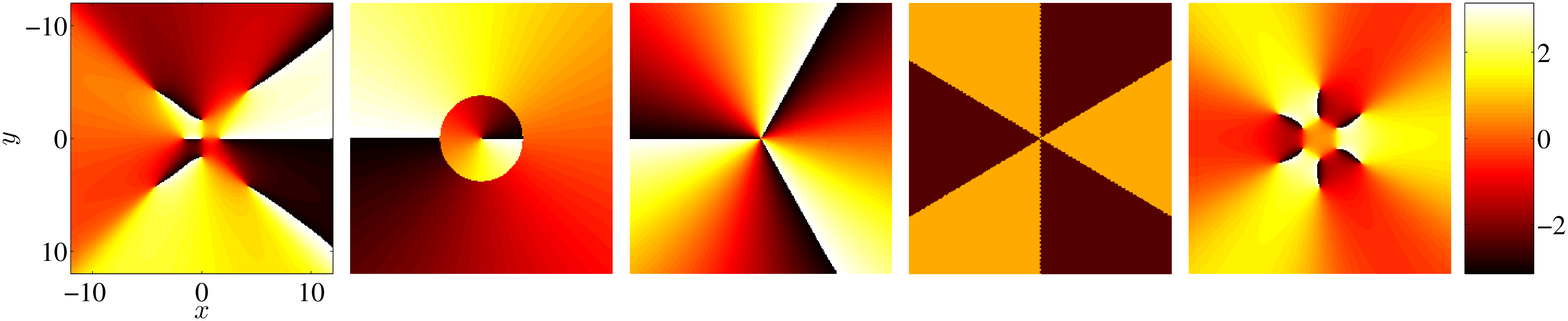} 
\includegraphics[height=.158\textheight, angle =0]{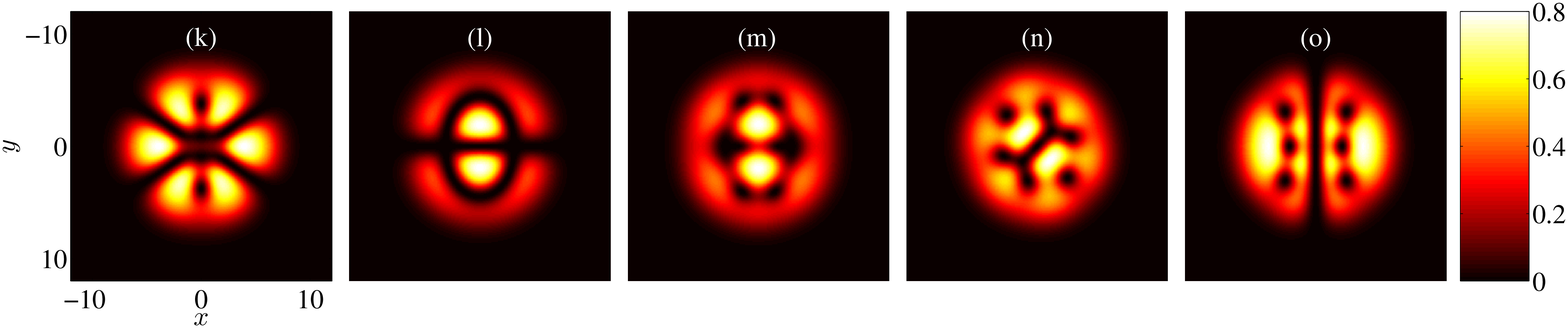} 
\includegraphics[height=.158\textheight, angle =0]{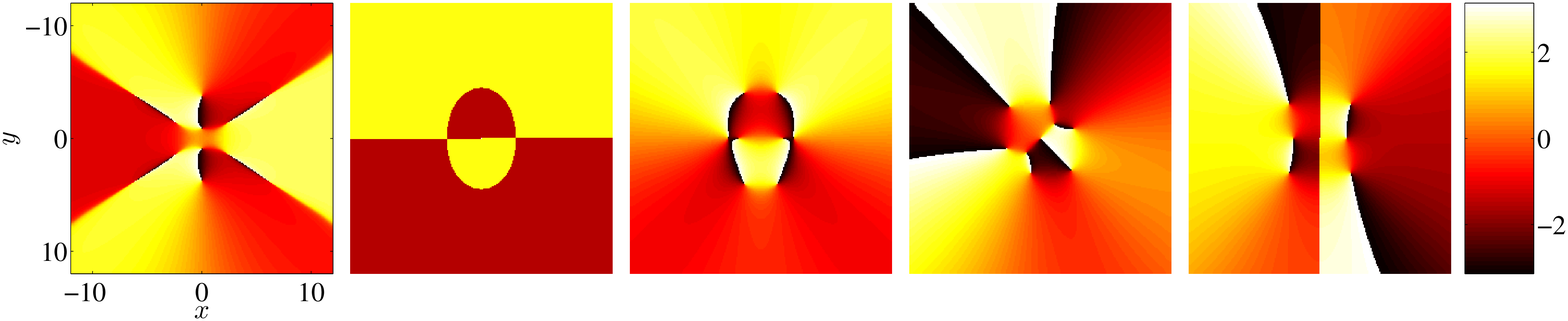} 
\end{center}
\caption{
(Color online) Same as Fig.~\ref{fig3} but for states
bifurcating from $\mu=4\Omega$. First, third and fifth 
as well as second, fourth and sixth rows correspond to 
plots of the density profiles and phases, respectively,
of the (a) $|3,0 \rangle_{(\textrm{c})}$ (at $\mu=1.28$) 
state, (b) $|2,1 \rangle_{(\textrm{c})}$ (at $\mu=1.29$) state, 
{(c) bifurcation emanating from (b) (at $\mu=1.3$)},
(d) nine vortex state ($\mu=1.23$), (e) $\ket{3,0}_{\textrm{c}}-i\ket{2,1}_{\textrm{c}}$
state ($\mu=1.32$), (f) different vortex necklace ($\mu=1.32$),
(g) vortex soliton ring ($\mu=1.21$), (h) vortex of topological 
charge $l=3$ ($\mu=1.32$), (i) soliton necklace associated with
$l=3$ mode ($\mu=1.23$), (j) vortex necklace ($\mu=1.32$), 
{(k) bifurcation emanating from (i) (at $\mu=1.32$)},
(l) $\Phi$-mode (discussed in the text) ($\mu=1.25$), 
{(m) bifurcation emanating from (l) (at $\mu=1.32$)}
(n) nine-vortex state with H-shape ($\mu=1.2$), and (o) six vortex
and dark soliton stripe ($\mu=1.32$),
respectively.}
\label{fig6}
\end{figure}

\begin{figure}[tbp]
\begin{center}
\includegraphics[height=.165\textheight, angle =0]{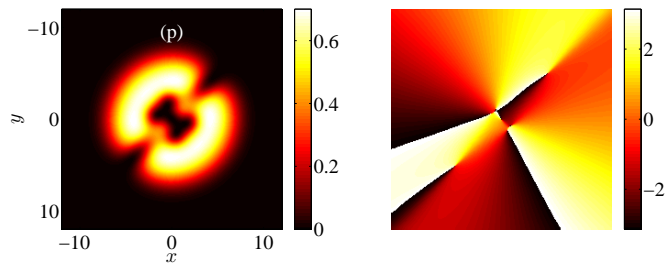}
\end{center}
\caption{
(Color online) Continuation of Fig.~\ref{fig6}.}
\label{fig6_supp_0}
\end{figure}

\begin{figure}[tbp]
\begin{center}
\includegraphics[height=.165\textheight, angle =0]{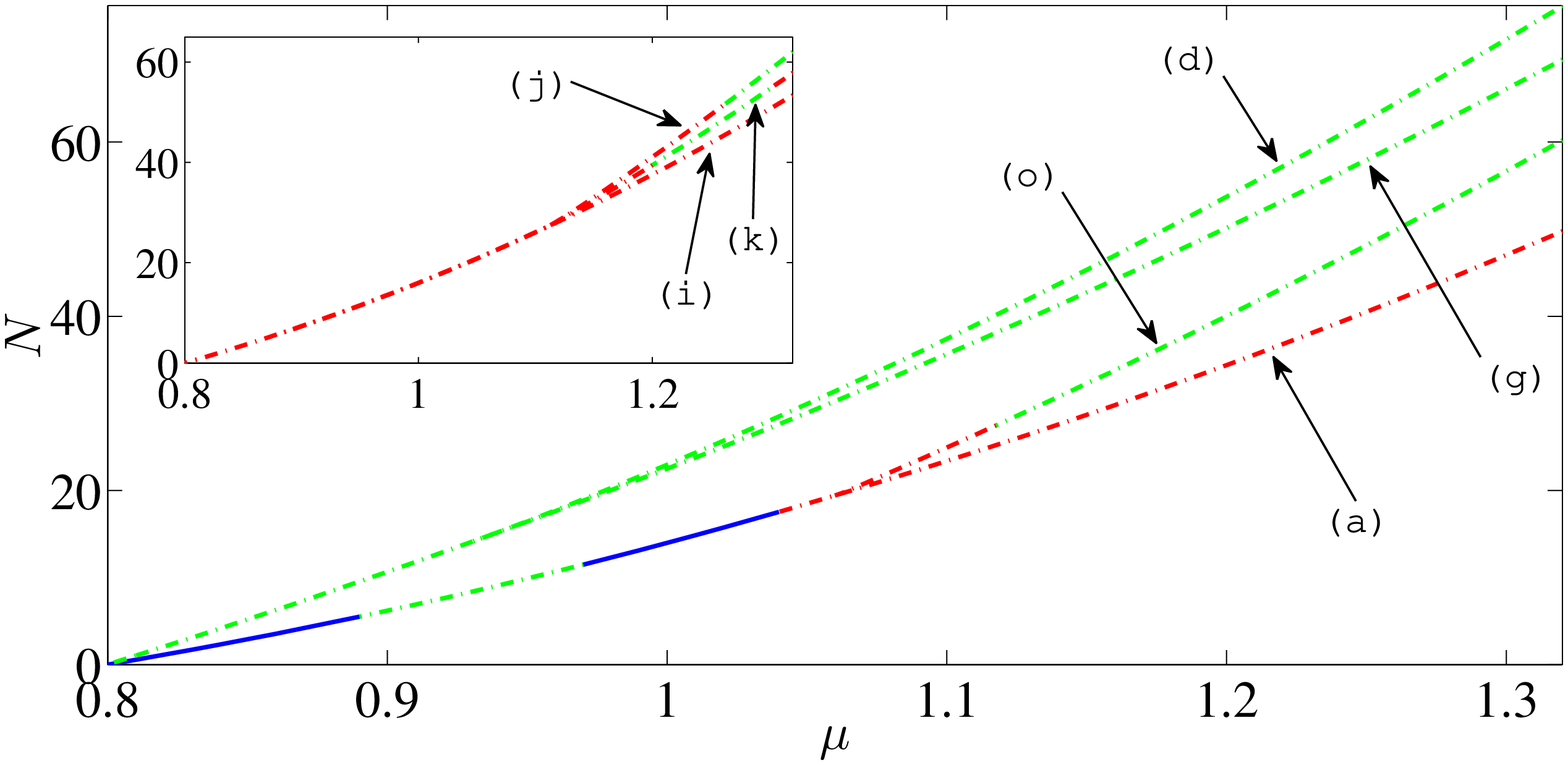}\includegraphics[height=.165\textheight, angle =0]{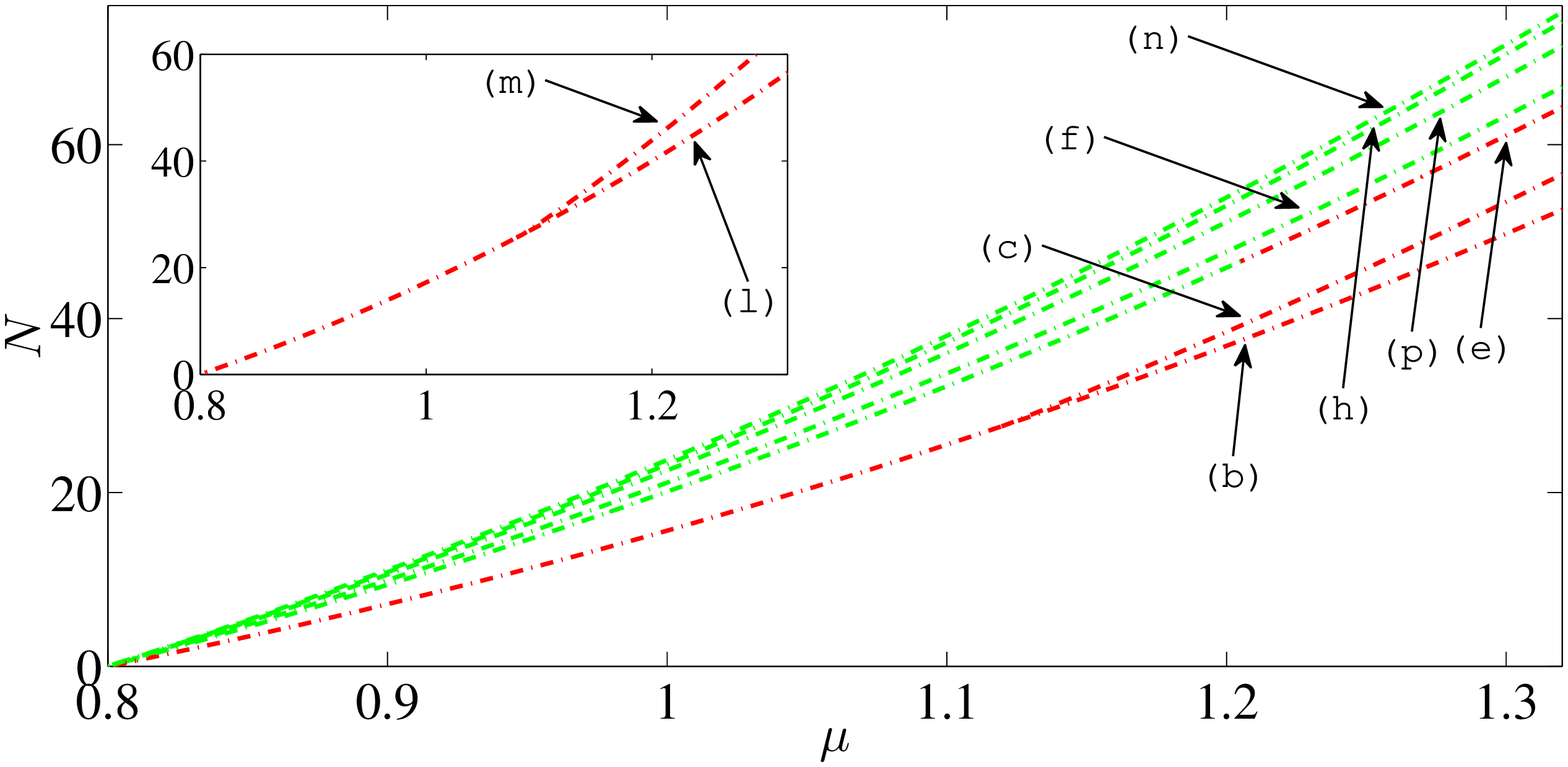}
\includegraphics[height=.165\textheight, angle =0]{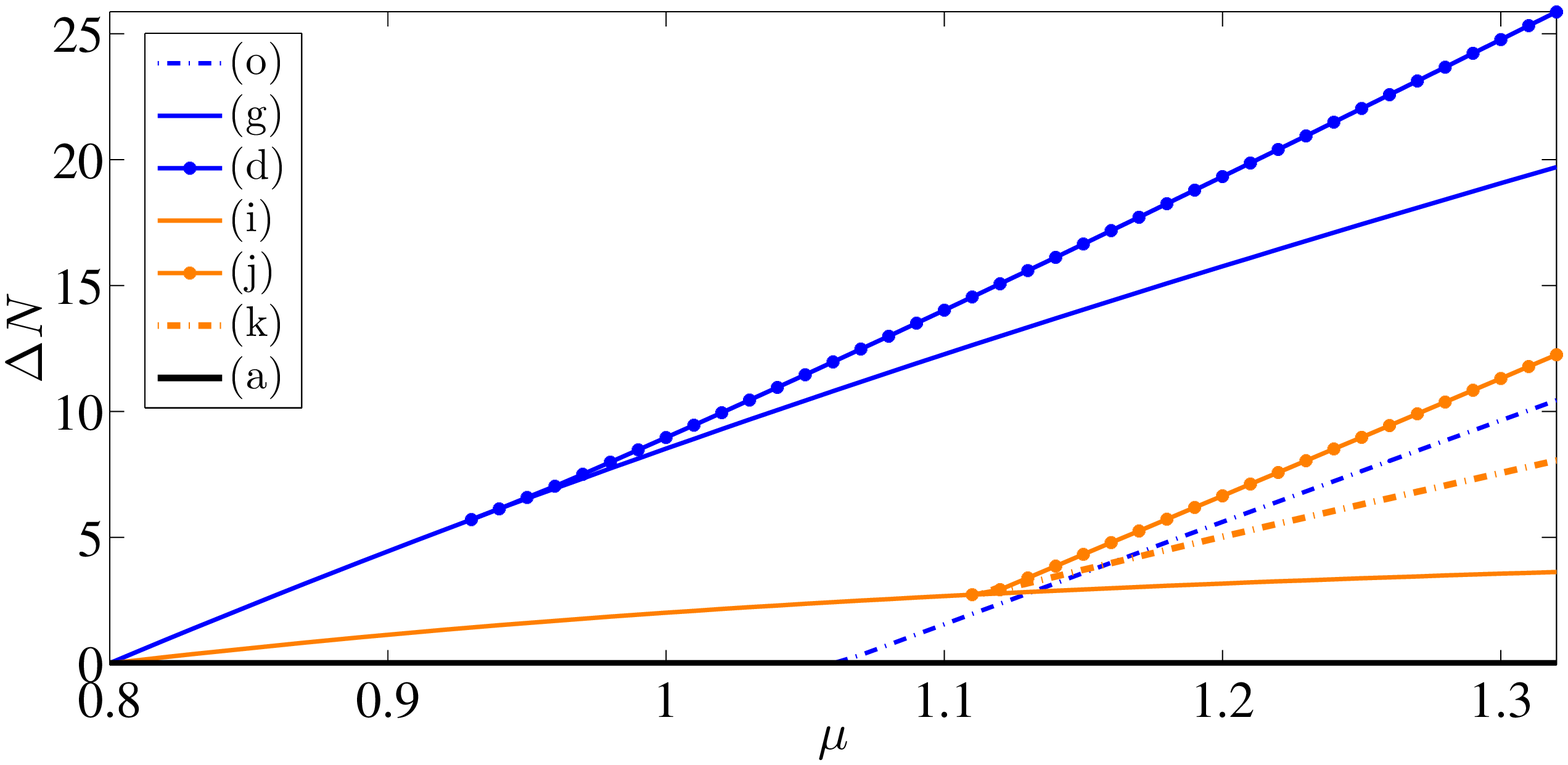}\includegraphics[height=.165\textheight, angle =0]{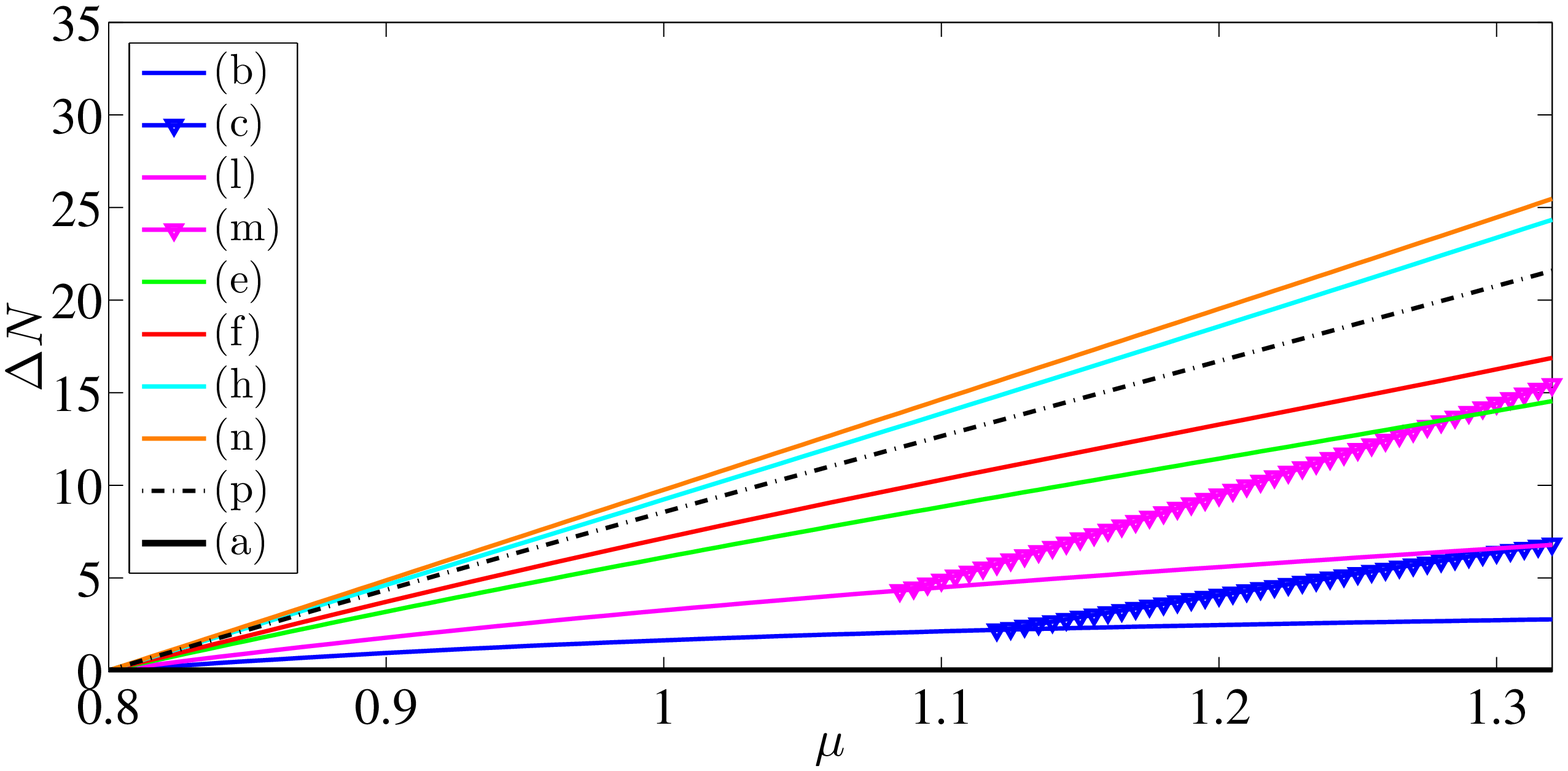}
\end{center}
\caption{
(Color online) Diagnostics as functions of $\mu$ for states
bifurcating from $\mu=4\Omega$: Top and bottom panels correspond
to the number of atoms and atom number difference with respect 
to the three dark soliton state, respectively. Bifurcations happen
at $\mu\approx0.93$, $\mu\approx1.06$, $\mu\approx1.09$, $\mu\approx1.11$,
and $\mu\approx1.12$ for Fig.~\ref{fig6}(d), Fig.~\ref{fig6}(o), Fig.~\ref{fig6}(m),
Figs.~\ref{fig6}(j)-(k), and Fig.~\ref{fig6}(c), respectively.}
\label{fig6_supp}
\end{figure}

We next examine the bifurcations emanating from $\mu
= 4\Omega$ with $n + m = 3$, plotted in Figs.~\ref{fig6}-\ref{fig6_supp}.
There are states that we can immediately recognize as
emerging out of the Cartesian linear limit, the $|3,0
\rangle_{(\textrm{c})}$ state of Fig.~\ref{fig6}(a) and the
$|2,1 \rangle_{(\textrm{c})}$ state of Fig.~\ref{fig6}(b).
The $|3,0 \rangle_{(\textrm{c})}$ state undergoes subsequent
bifurcations leading to real eigenvalues and symmetry-breaking
destabilizations,  e.g., the six vortex and dark
soliton stripe state of Fig.~\ref{fig6}(o)
(see, also,~Fig.~6 of \cite{middel2}).
This pitchfork bifurcation, due to the admixture of
$|3,0 \rangle_{(\textrm{c})}$ with $|1,3 \rangle_{(\textrm{c})}$
with a $\pi/2$ shift (see the relevant theory of~\cite{middel2}),
leads to the destabilization of the former branch.
Furthermore, the $|2,1 \rangle_{(\textrm{c})}$ mode gives
  birth to a state having two dark solitons and two vortices
  -through admixture with $|4,0 \rangle_{(\textrm{c})}$ again with
  a $\pi/2$ phase shift-. The bifurcating
state appears to be exponentially unstable, inheriting the
instability of its ``parent branch'', over the full parametric
interval of $\mu$ and is depicted in Fig.~\ref{fig6}(c).
In addition, the algorithm has discovered states that can be 
naturally expressed as linear combinations of the Cartesian 
eigenstates. The state depicted in Fig.~\ref{fig6}(d) can be 
expressed as $|3,0\rangle_{(\textrm{c})} + i |0,3 \rangle_{(\textrm{c})}$
in the linear limit and represents a lattice of 9 vortices of 
alternating charge and was previously characterized via algebraic
conditions~\cite{generating}. This solution appears to be oscillatorily
unstable throughout our computations. 

The states depicted in Fig.~\ref{fig6}(e) and (f) can be approximated
by the linear combinations $\ket{3,0}_{(\textrm{c})}-i\ket{2,1}_{(\textrm{c})}$ 
and $\ket{2,1}_{(\textrm{c})}+i\ket{1,2}_{(\textrm{c})}$,
respectively. To the best of our knowledge, these solutions have 
not been previously identified (cf. Fig. 6 of~\cite{middel2}),
yet adhere to the formulation whereby more complex nonlinear 
solutions generalize linear combinations of simpler linear eigenstates.
Furthermore, in a manner reminiscent of Fig.~\ref{fig3}(g), such solutions
feature a pattern of vortices (along a line or on a cross, respectively)
close to the center of the trap and another one further away. Both 
branches appear to be oscillatorily unstable, except for an interval 
where the former branch also develops exponential instabilities
associated with real eigenvalues for $\mu \gtrsim 1.21$.

There are also states such as the ring-vortex state of
Fig.~\ref{fig6}(g) (previously examined, e.g., in~\cite{carr,herring})
and the well-known triple charge vortex of Fig.~\ref{fig6}(h) that
are best described in polar notation, $\ket{1,1}_{(\textrm{p})}$
and $\ket{0,3}_{(\textrm{p})}$ respectively. Both branches are
oscillatorily unstable.  In the terminology of \cite{toddrotating},
Fig.~\ref{fig6}(i) depicts a multi-pole, a real solution of the form
$q_{0,3}(r) \cos(3\theta)$, whereas a complex combination of $q_{1,1}(r)$
with $q_{0,3}(r) \sin(3 \theta)$ will produce the vortex necklace 
of Fig.~\ref{fig6}(j). Furthermore, the state of Fig.~\ref{fig6}(i)
  undergoes a symmetry-breaking bifurcation
  --through an admixture with $|0,4 \rangle_{(\textrm{c})}$ with a
  $\pi/2$ phase shift--
  leading to the state of 
  Fig.~\ref{fig6}(k). The latter possesses two isolated vortices
  (two more appear ``hidden'' in the region of vanishing density
  and can be discerned in the associated phase plot). It is
  exponentially
  unstable, inheriting the instability of its multi-pole parent
  branch. This is with the exception of a narrow window (in $\mu$) where the
oscillatory instability dominates.
The branch of Fig.~\ref{fig6}(l) is associated with $q_{1,1}(r) \cos(\theta)$
i.e., another type of multi-pole (or $\ket{2,1}_{(\textrm{c})}+\ket{0,3}_{(\textrm{c})}$
in the Cartesian format) and is referred to as a $\Phi$ soliton in
the terminology of~\cite{brand}. However, this state
  becomes subject  to instability similar to that leading
  to branch Fig.~\ref{fig3}(c), associated with a hexagonal mode.
  As a result, the daughter branch of Fig.~\ref{fig6}(m) emerges
  possessing four isolated vortices at the top and bottom and two
  additional charge $2$ vortices along the x-axis.

The bifurcation diagram in Fig.~\ref{fig6_supp} sheds light on
the potential stability of the branches, as well as on the bifurcations
that arise. 
More concretely, we observe that almost all the branches are now 
dynamically unstable, a feature that is not surprising given the highly excited 
nature of the states. Nevertheless, the branch $|3,0 \rangle_{(\textrm{c})}$
of Fig.~\ref{fig6}(a)
possesses intervals of stability. The top left panel of Fig.~\ref{fig6_supp} sheds 
light on relevant bifurcations including the fact that the 9-vortex state of
Fig.~\ref{fig6}(d) emerges from the ring-vortex branch of  Fig.~\ref{fig6}(g),
while the vortex necklace of Fig.~\ref{fig6}(j) {and the state of
Fig.~\ref{fig6}(k)} emerge as bifurcations from the soliton necklace of Fig.~\ref{fig6}(i). 
Finally, the 6-vortex plus soliton state of Fig.~\ref{fig6}(o) emerges as a 
bifurcation from the three-dark-soliton stripe state of Fig.~\ref{fig6}(a).

Arguably, of particular note are the complex patterns
of Figs.~\ref{fig6}(n) and \ref{fig6_supp_0}(p), both emerging from
the linear limit. The state of Fig.~\ref{fig6}(n) represents a vortex 
pattern involving 9 vortices (and total charge 3) where seven of 
the vortices are involved in an elaborate vortical H-shape, while 
the other two are distinct. This state (and the one shown in Fig.~\ref{fig6_supp_0}(p))
can be classified as emerging from a complex (both literally and
figuratively) combination at the linear limit. To unveil
the relevant superposition, we project the mode --in the immediate
vicinity of the linear limit--
onto the fundamental modes $\ket{m,n}_{\textrm{(c)}}$ (again, with $m+n=3$). 
This gives rise to
the combination: $\alpha(\ket{3,0}_{\textrm{(c)}}+i\,\ket{2,1}_{\textrm{(c)}})%
+i\alpha^{\ast}(\ket{0,3}_{\textrm{(c)}}-i\ket{1,2}_{\textrm{(c)}})$
with $\alpha$ being a suitable complex prefactor; in the
case we considered, a typical value of $\alpha$ was found
as: $\alpha\approx -0.15 + 0.5\,i$. In a similar vein, the projection of 
the state of Fig.~\ref{fig6_supp_0}(p) onto the fundamental modes suggests
the combination: $\alpha(\ket{3,0}_{\textrm{(c)}}-i\,\ket{1,2}_{\textrm{(c)}})%
+\alpha^{\ast}(\ket{0,3}_{\textrm{(c)}}+i\ket{2,1}_{\textrm{(c)}})$
with $\alpha\approx 0.14 + 0.04\,i$ in this case. To the best of our knowledge,
such patterns have also not been previously identified and are direct by-products
of the use of deflated continuation.


These bifurcations are best understood by plotting the diagnostic 
$\Delta N$, where the base solution is taken to be the three (planar) 
dark soliton branch $|3,0\rangle_{(\textrm{c})}$. The
resulting bifurcation diagram is presented in the bottom panels of Fig.~\ref{fig6_supp}.

\subsection{Bifurcations from $\mu = 5\Omega$}
In the case of $\mu = 5 \Omega$, the wealth of relevant 
states is even greater. We first discuss the states that
are naturally expressed in Cartesian format.

\begin{figure}[tbp]
\begin{center}
\includegraphics[height=.158\textheight, angle =0]{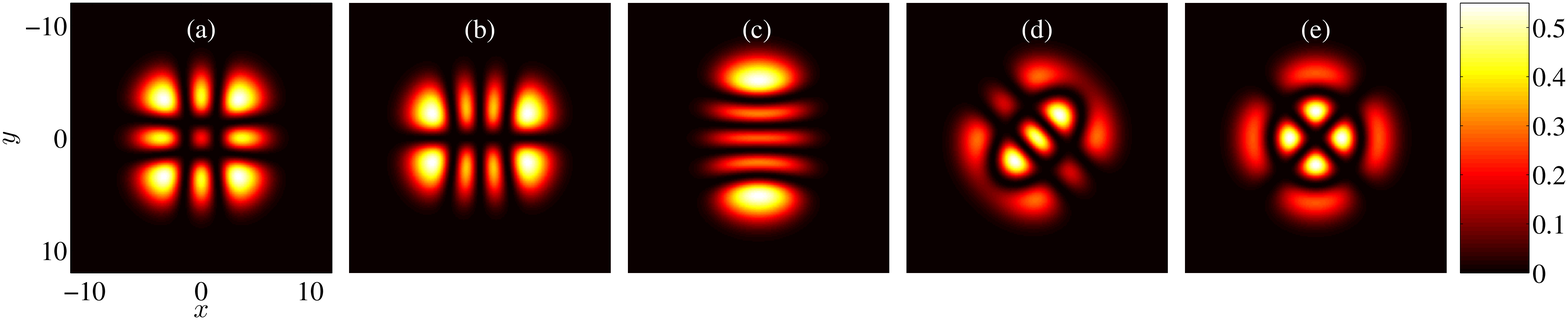} 
\includegraphics[height=.158\textheight, angle =0]{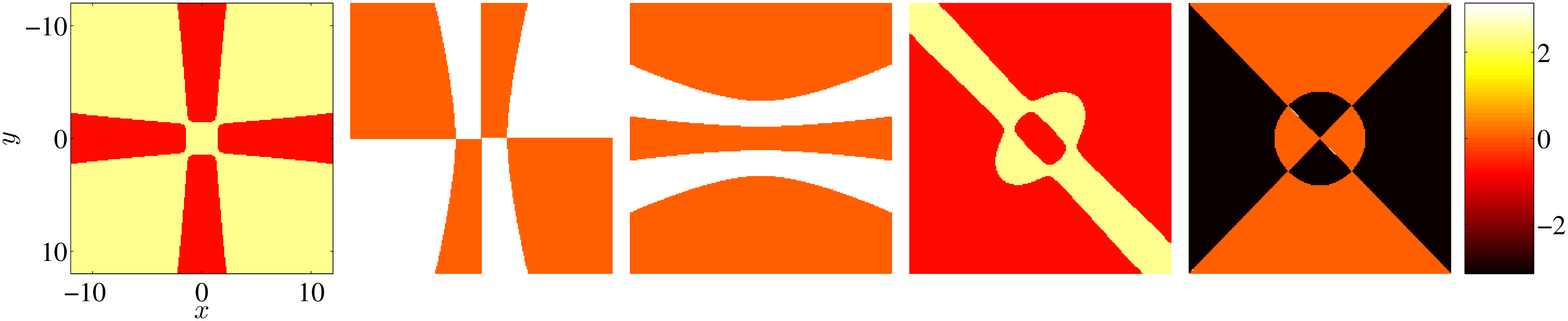}
\includegraphics[height=.158\textheight, angle =0]{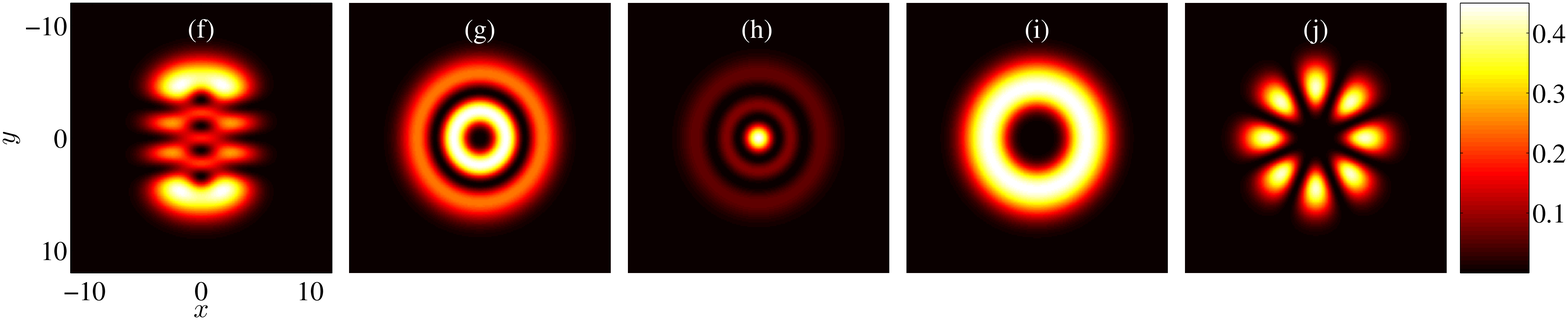} 
\includegraphics[height=.158\textheight, angle =0]{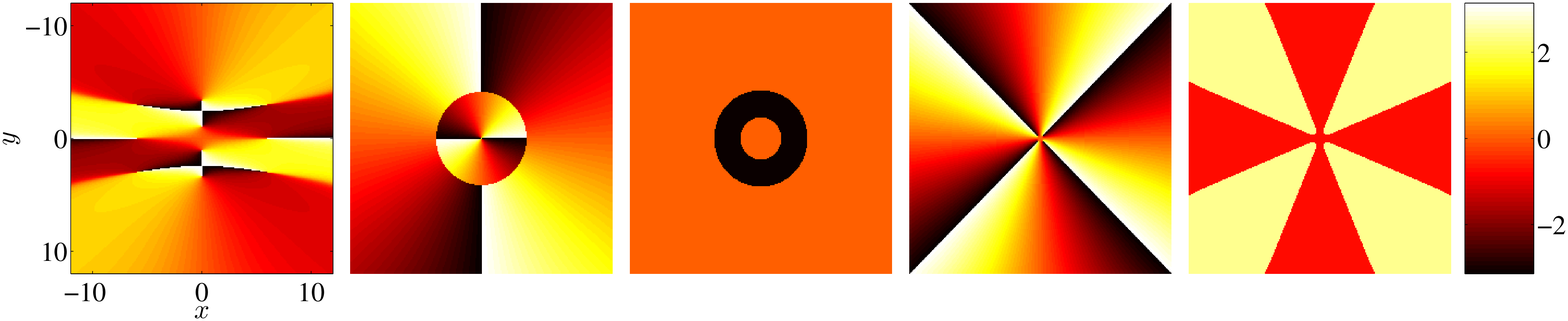} 
\includegraphics[height=.158\textheight, angle =0]{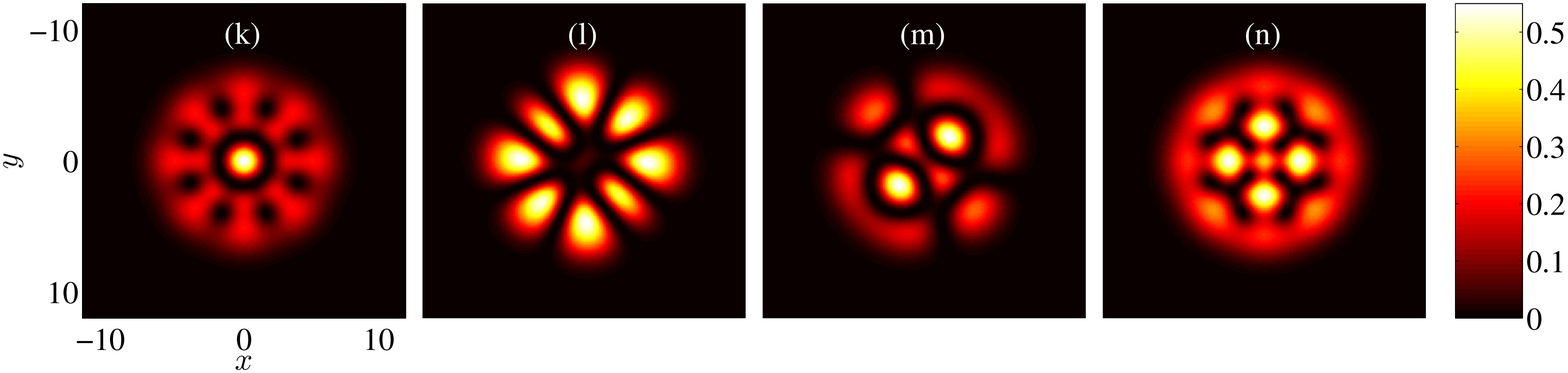} 
\includegraphics[height=.158\textheight, angle =0]{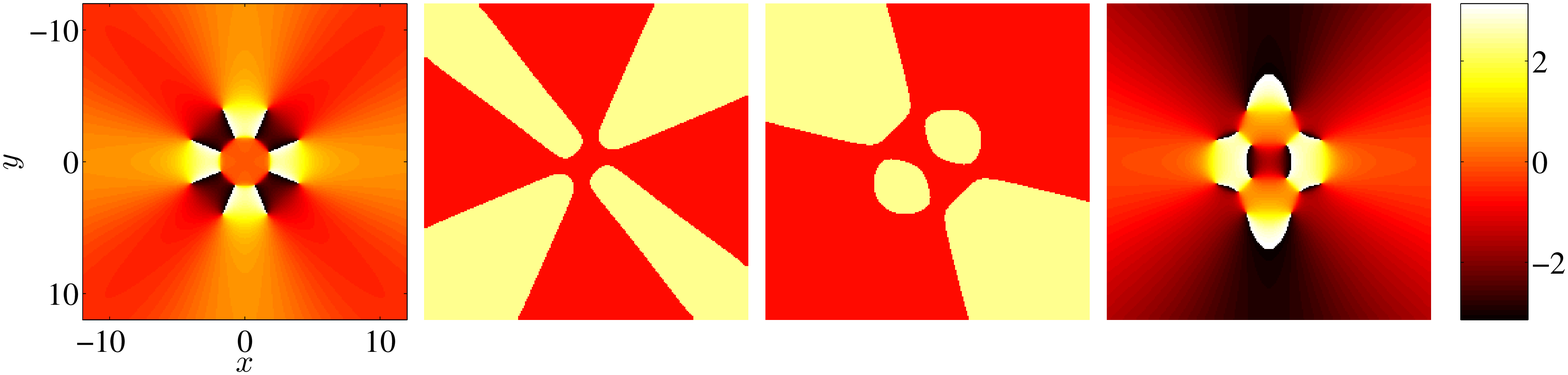}
\end{center}
\caption{
(Color online) Same as Fig.~\ref{fig6} but for states bifurcating 
from $\mu=5\Omega$. Panels (a)-(c) represent the $\ket{2,2}_{(\textrm{c})}$ ($\mu=1.32$),
$\ket{3,1}_{(\textrm{c})}$ ($\mu=1.32$), and $\ket{0,4}_{(\textrm{c})}$ ($\mu=1.32$)
states, respectively, while (d) shows a double $\Phi$ state
corresponding to $|4,0\rangle_{(\textrm{c})}+|2,2 \rangle_{(\textrm{c})}$
and (e) is associated with
$|4,0 \rangle_{(\textrm{c})}-|0,4 \rangle_{(\textrm{c})}$ ($\mu=1.23$).
Panel (f) represents the $\ket{4,0}_{(\textrm{c})}-i\ket{3,1}_{(\textrm{c})}$
($\mu=1.25$) state. Furthermore, panel (g) is a ring dark soliton ($\mu=1.23$) 
harboring an $l=2$ vortex at its center; panel (h) is a double ring
($\mu=1.08$); panel (i) represents the $\ket{0,4}_{(\textrm{p})}$ state ($\mu=1.32$). 
Panel (j) is a multi-pole ($\mu=1.25$) (a purely real solution) and (k) 
corresponds to the $\ket{2,0}_{(\textrm{p})}+iq_{0,4}\cos{(4\theta)}$ state
($\mu=1.15$) and is a so-called vortex necklace. Panel (l) represents a soliton
necklace (real profile) state ($\mu=1.32$) (see text for details). The double
stripe+ring ($\mu=1.23$) of panel (e) has also a similar curved/bent variant in
panel (m) ($\mu=1.2$). Finally, panel (n) is a vortex necklace ($\mu=1.23$) in 
the terminology of~\cite{toddrotating}.
}
\label{fig7}
\end{figure}

Figs.~\ref{fig7}(a), (b) and (c) are associated with the
$\ket{2,2}_{(\textrm{c})}$, $|3,1 \rangle_{(\textrm{c})}$
and $\ket{0,4}_{(\textrm{c})}$ states respectively. All of
these branches are potentially subject to exponential instabilities except
for the branch (c) which bears oscillatory instabilities for 
$\mu\lesssim 1.235$ and for $\mu\lesssim 1.06$ the waveform
is stabilized. Additional states can be produced from linear 
combinations of the Cartesian eigenstates. In particular, the
states of Figs.~\ref{fig7}(d) and (e) can be approximated in 
their linear limit as $|4,0\rangle_{(\textrm{c})}+|2,2 \rangle_{(\textrm{c})}$
and $|4,0 \rangle_{(\textrm{c})}-|0,4 \rangle_{(\textrm{c})}$
respectively. The former can be thought of as a double $\Phi$ 
solution. Both solutions appear to be exponentially unstable in
our computations. In addition, the branch of Fig.~\ref{fig7}(f) 
can be characterized by a linear combination of the form of 
$|4,0 \rangle_{(\textrm{c})}-i|3,1 \rangle_{(\textrm{c})}$.
This branch appears to be oscillatorily unstable in our stability 
analysis.

Other states are more naturally classified in the polar 
representation. For instance, Fig.~\ref{fig7}(g) corresponds
to the mode $\ket{1,2}_{(\textrm{p})}$, 
associated with an oscillatorily unstable double vortex at the origin,
bearing also a ring dark soliton in the periphery. Similarly, 
Fig.~\ref{fig7}(h) can be represented as $\ket{2,0}_{(\textrm{p})}$
and corresponds to a highly unstable double ring configuration. 
The vortex of charge four depicted in Fig.~\ref{fig7}(i) can be 
represented as $\ket{0,4}_{(\textrm{p})}$ and is oscillatorily unstable;
all of these states could be identified in the polar decomposition 
of~\cite{carr,herring}. Other states are generalizations of ones that
we identified in Fig.~\ref{fig6}. Fig.~\ref{fig7}(j) depicts a multi-pole;
this is described by $q_{0,4}(r) \cos(4 \theta)$ in its linear limit and
is subject to exponential instabilities. 

\begin{figure}[tbp]
\begin{center}
\includegraphics[height=.19\textheight, angle =0]{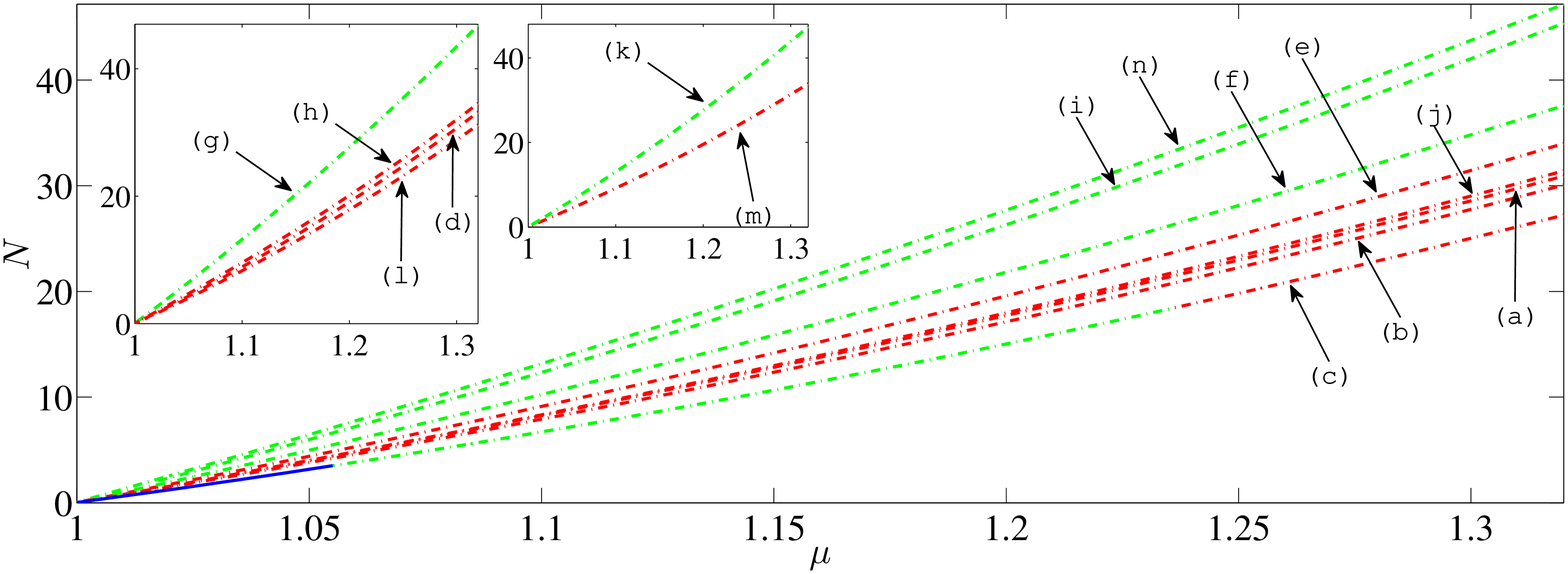}
\end{center}
\caption{
(Color online) The number of atoms $N$ as a function of $\mu$
and for states bifurcating from $\mu=5\Omega$.  
}
\label{fig7_supp}
\end{figure}

Additional states can be
produced from linear combinations of the polar eigenstates. 
Fig.~\ref{fig7}(k) can be represented by a complex superposition of 
the double ring configuration and the soliton necklace; it corresponds to
the $\ket{2,0}_{(\textrm{p})}+iq_{0,4}(r) \cos(4 \theta)$ state which is
oscillatorily unstable. This is a canonical example of a vortex necklace
in the terminology of~\cite{toddrotating} (and as can also be seen in 
the figure; cf. Fig. 1(d) of~\cite{toddrotating}). Other states are more 
difficult to classify, although it is still plausible to classify
them by using ``exotic'' combinations of Cartesian and polar eigenmodes. 
For instance, Fig.~\ref{fig7}(l) depicts a solution that can be described
by $q_{0,4}(r) \cos(4 \theta)-2\ket{2,2}_{(\textrm{c})}+\ket{0,4}_{(\textrm{c})}$
and is subject to exponential instabilities. This real solution is a soliton
necklace in the terminology of~\cite{toddrotating}. Fig.~\ref{fig7}(m)
corresponds to the ``curved'' variant of Fig.~\ref{fig7}(e) (subject to 
an exponential instability, as well). 
The vortex necklace of Fig.~\ref{fig7}(n)
appears to be subject only to oscillatory instabilities. This is an elaborate
pattern, once again revealed by the technique of deflation, that may be thought of as consisting of 4 vortex triplets
(in a Y shape) each of which adds a charge of $1$ to the total charge
of $4$ within the structure. In terms of stability, we observe in Fig.~\ref{fig7_supp}
that the branch (c) is linearly stable near the linear limit.


\subsection{Bifurcations from $\mu = 6\Omega$}
\begin{figure}[tbp]
\begin{center}
\includegraphics[height=.158\textheight, angle =0]{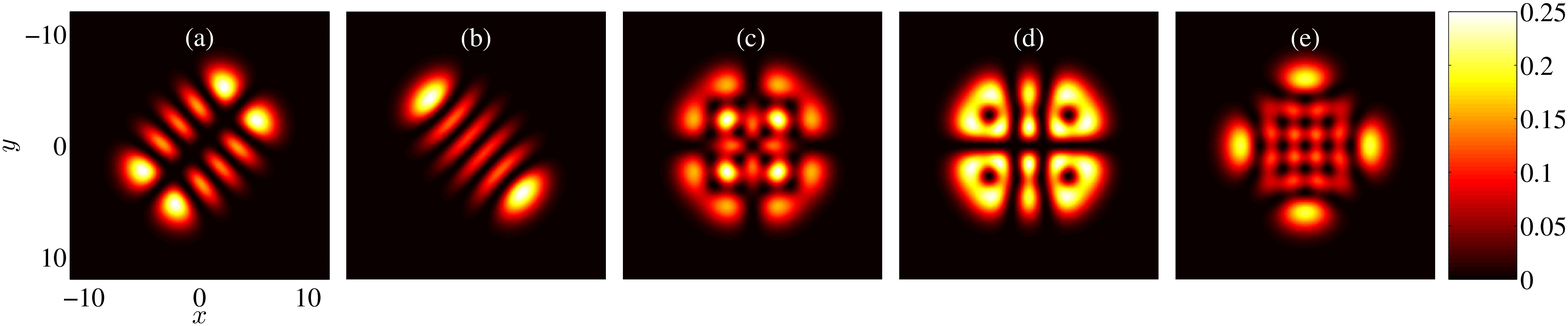} 
\includegraphics[height=.158\textheight, angle =0]{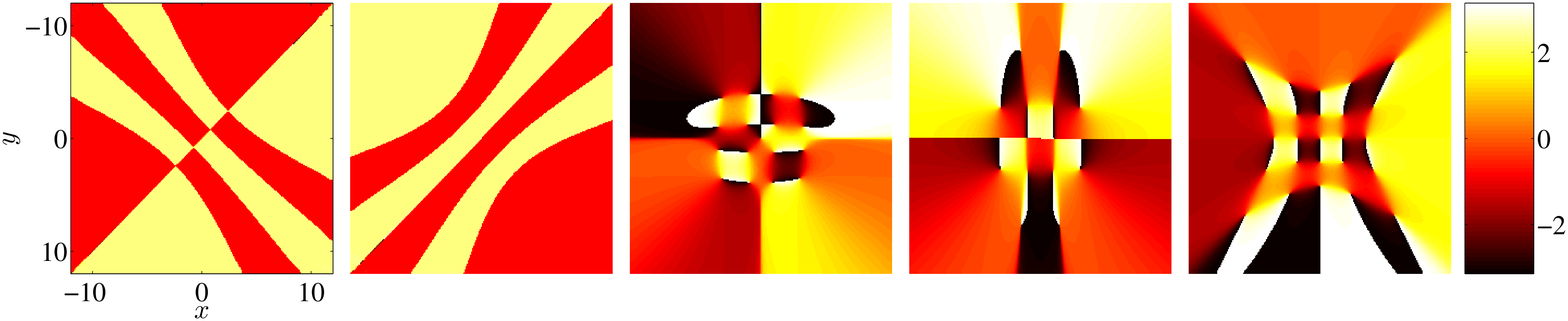}
\includegraphics[height=.158\textheight, angle =0]{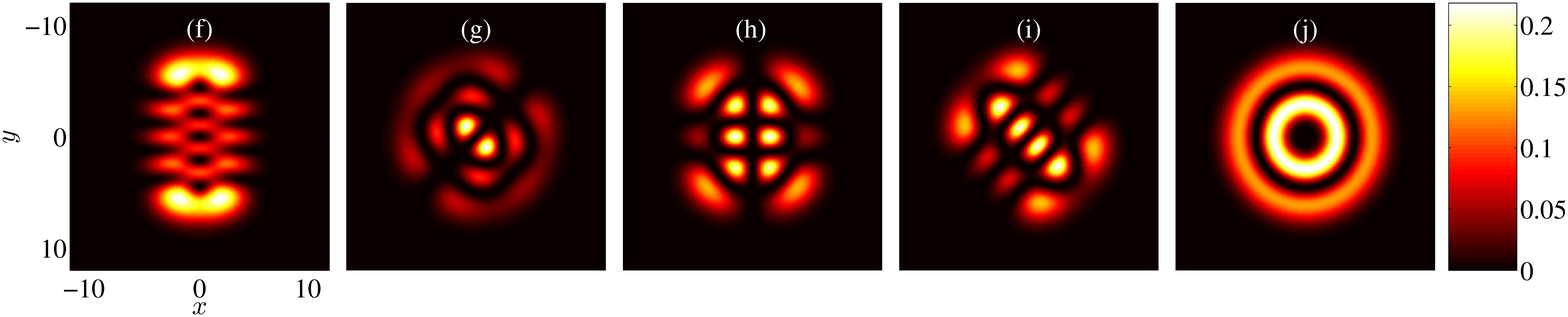} 
\includegraphics[height=.158\textheight, angle =0]{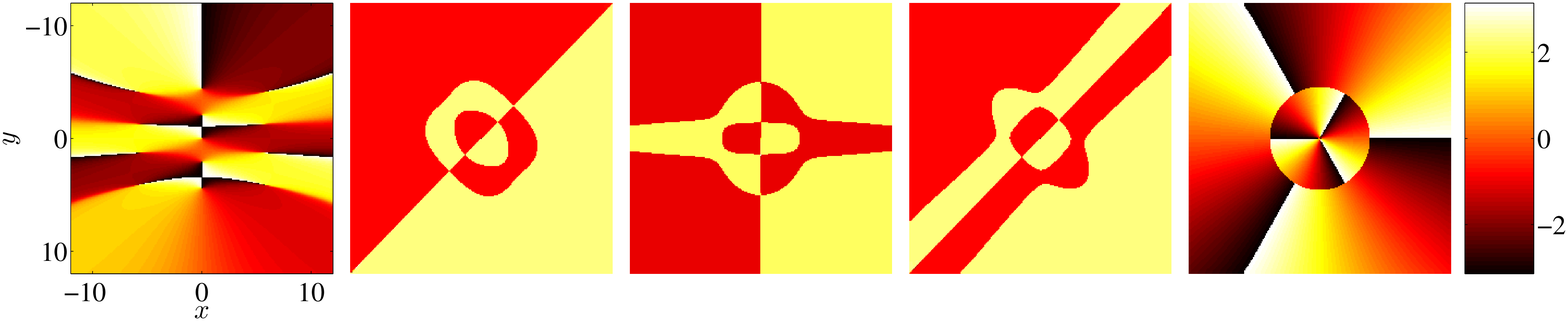}
\includegraphics[height=.158\textheight, angle =0]{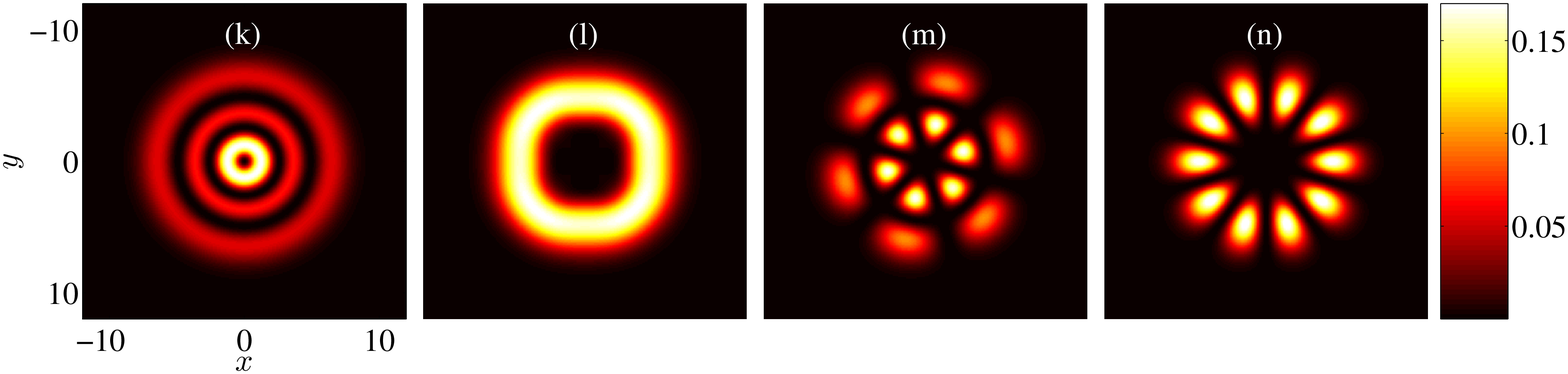} 
\includegraphics[height=.158\textheight, angle =0]{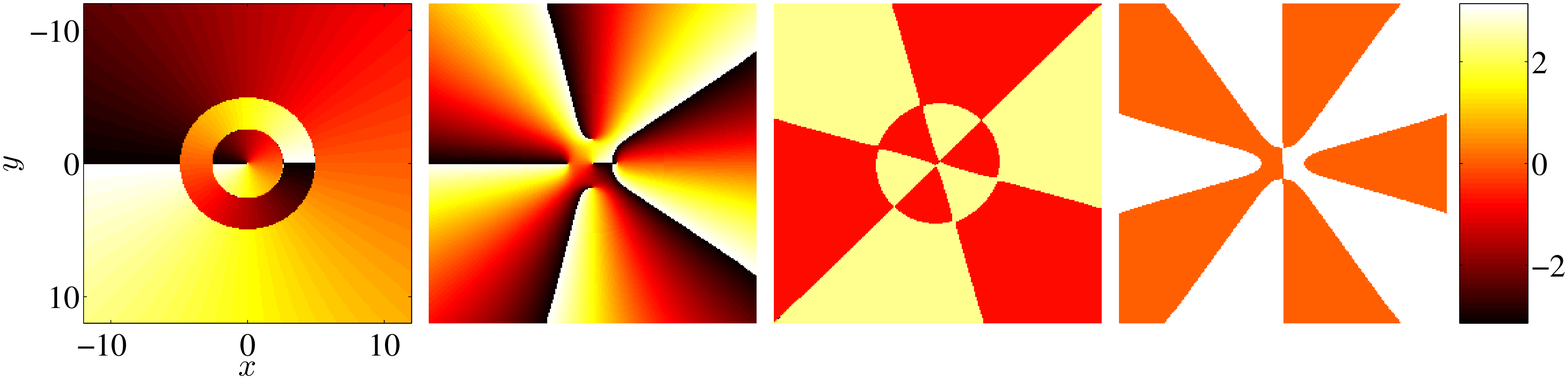}
\end{center}
\caption{
(Color online) 
Same as Fig.~\ref{fig7} but for states bifurcating from 
$\mu=6\Omega$.  First, third and fifth as well as second, 
fourth and sixth rows correspond to plots of the density 
profiles and phases, respectively, of the (a) $\ket{4,1}_{(\textrm{c})}$ state ($\mu=1.3$), 
(b) $\ket{0,5}_{(\textrm{c})}$ state ($\mu=1.29$), (c) $\ket{1,4}_{(\textrm{c})}+i\ket{4,1}_{(\textrm{c})}$ state ($\mu=1.28$), 
(d) $\ket{3,2}_{(\textrm{c})}+i\ket{1,4}_{(\textrm{c})}$ ($\mu=1.32$),
(e) $\ket{5,0}_{(\textrm{c})}-i\ket{0,5}_{(\textrm{c})}$ state ($\mu=1.295$), 
(f) $\ket{5,0}_{(\textrm{c})}-i\ket{4,1}_{(\textrm{c})}$ state ($\mu=1.31$),
(g) $\ket{5,0}_{(\textrm{c})}-\ket{0,5}_{(\textrm{c})}$ state ($\mu=1.25$),
(h) $\ket{3,2}_{(\textrm{c})}+\ket{1,4}_{(\textrm{c})}$ state ($\mu=1.29$),
(i) triple $\Phi$ solution ($\mu=1.29$), 
(j) ring-dark soliton ($\mu=1.32$), 
(k) double ring configuration ($\mu=1.262$),
(l) vortex with charge $l=5$ ($\mu=1.32$), 
(m) double solitonic necklace ($\mu=1.27$), 
and (n) solitonic necklace ($\mu=1.297$), respectively.
}
\label{fig8}
\end{figure}

Finally, we examine an even more ``exotic'' set of
solutions, the states bifurcating from the linear limit 
at $\mu=6 \Omega$. Here too we observe that most of the
states are oscillatorily unstable for sufficiently large 
$\mu$, with the exception of the branches of the solutions
depicted in Figs.~\ref{fig8}(a), (g)--(i), (m) and (n) that
seem to be dominated by an exponential instability. 

Many of the
bifurcating states can be naturally classified in Cartesian
form. Fig.~\ref{fig8}(a) represents the $|4,1 \rangle_{(\textrm{c})}$
state, while Fig.~\ref{fig8}(b) represents the $|5,0 \rangle_{(\textrm{c})}$
state. The states depicted in Figs.~\ref{fig8}(c), (d), (e)-(g), 
(h) and (i) can be approximated by linear combinations of the Cartesian
eigenstates; they can approximated as $\ket{1,4}_{(\textrm{c})}+i\ket{4,1}_{(\textrm{c})}$,
$\ket{3,2}_{(\textrm{c})}+i\ket{1,4}_{(\textrm{c})}$, 
$\ket{5,0}_{(\textrm{c})}-i\ket{0,5}_{(\textrm{c})}$,
$\ket{5,0}_{(\textrm{c})}-i\ket{4,1}_{(\textrm{c})}$,
$\ket{5,0}_{(\textrm{c})}-\ket{0,5}_{(\textrm{c})}$ 
(the ``double loop'' $\Phi$ solution with the two radii being concentric),
$\ket{3,2}_{(\textrm{c})}+\ket{1,4}_{(\textrm{c})}$,
$\ket{5,0}_{(\textrm{c})}+\ket{3,2}_{(\textrm{c})}$ (triple $\Phi$ solution),
respectively. It should be noted that \emph{none} of these
solutions has previously appeared in a systematic fashion
in the literature, to the best of our knowledge.

Others are more naturally described using the polar
decomposition. Fig.~\ref{fig8}(j) depicts a ring dark
soliton with a vortex of charge 3 at its center, i.e.,
a $\ket{1,3}_{(\textrm{p})}$. Fig.~\ref{fig8}(k) depicts
a double ring configuration with a vortex of charge 1 
represented by $\ket{2,1}_{(\textrm{p})}$. Fig.~\ref{fig8}(l)
depicts the vortex of charge 5, i.e., $\ket{0,5}_{(\textrm{p})}$.
Fig.~\ref{fig8}(m) depicts a double solitonic necklace.
Finally, the multi-pole of Fig.~\ref{fig8}(n) can 
be represented in the polar decomposition as $q_{0,5}\cos{(5\theta)}$. 

It is already clear at this stage that the full
classification of the pertinent solutions becomes
an extremely cumbersome task. This motivates the
application of methods from bifurcation analysis to
yield a comprehensive perspective that might not otherwise
be available. In this case, deflated continuation has
made it possible to unravel a wide variety of branches
that had not been previously identified.

\begin{figure}[tbp]
\begin{center}
\includegraphics[height=.19\textheight, angle =0]{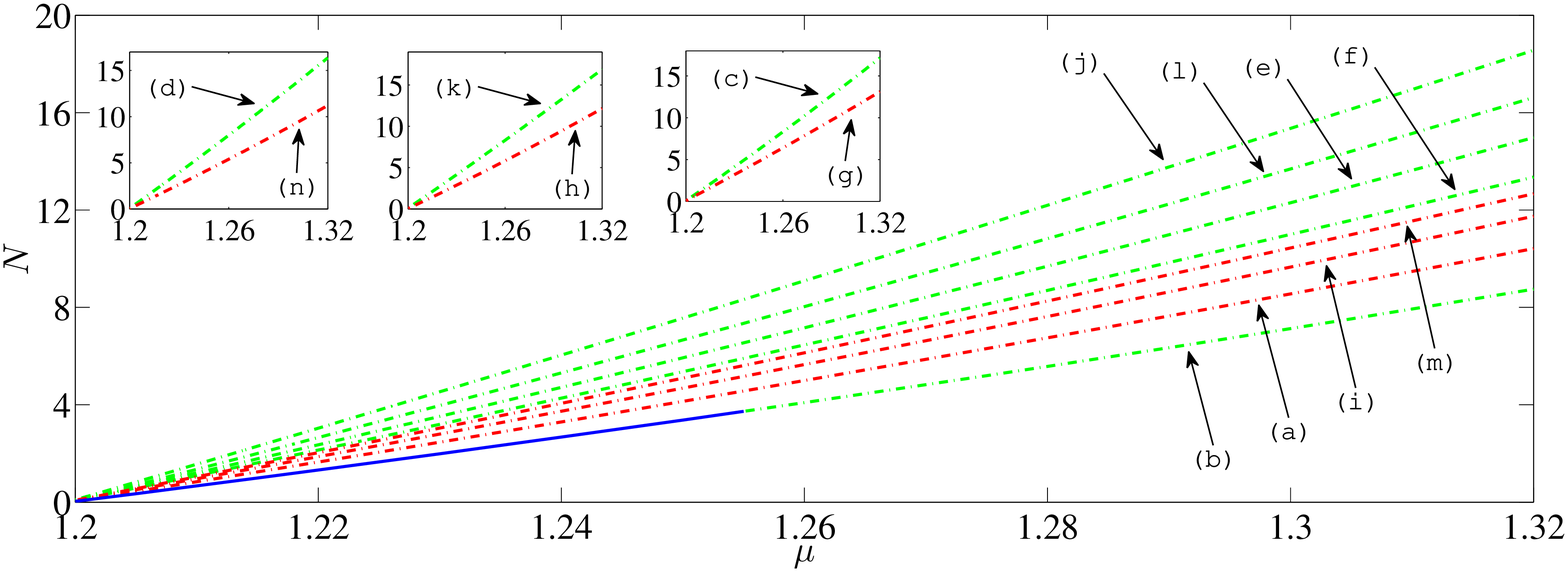}
\end{center}
\caption{
(Color online) Same as Fig.~\ref{fig7_supp} but for states 
bifurcating from $\mu=6\Omega$. The number of atoms $N$ as 
a function of $\mu$. 
}
\label{fig8_supp}
\end{figure}

As regards the stability of the branches, 
Fig.~\ref{fig8_supp} suggests that the branch in Fig.~\ref{fig8}(b),
i.e., the  $|5,0 \rangle_{(\textrm{c})}$ state is stable
over a narrow parametric interval in $\mu$.


\section{Concluding remarks and future challenges} \label{sec:conclusion}

In this work we applied deflated continuation to the
analysis of the solutions of a Bose-Einstein condensate in a
two-dimensional isotropic parabolic trap. This problem
has a well-understood linear limit in which the eigenmodes
can be identified in closed analytical form, offering
guidance for what to expect within the nonlinear regime.
There has been a wide range of publications on this system
that have revealed a broad spectrum of nonlinear excitations
and their  stability characteristics. 
However, the number of solutions and their complexity
increase significantly as the chemical potential increases
and analytical calculations become rather tedious.
Deflated continuation is therefore extremely useful in yielding insight
into the problem.
We have identified branches (from the linear limit) and
configurations bearing complex vortex patterns
as genuine (although often unstable) solutions of the system.
We have also identified the bifurcations involving such
configurations including, to give but one example, the
elaborate multi-vortex
patterns that arise from the symmetry breaking
e.g, for $l=2$ (Figs.~\ref{fig5}(d) and (e)), $l=3$ (Figs.~\ref{fig3}(c)), 
and so on.

The use of deflated continuation enabled us to identify a
wide range of branches that had not been previously numerically
constructed or continued.
Nevertheless, it should be highlighted that {\it not all}
branches obtained herein were identified using this method alone;
deflated continuation is not guaranteed to find all solutions
to a given nonlinear problem.
In particular, the solutions depicted in Figs.~\ref{fig4}(d),~\ref{fig5}(c), 
\ref{fig6}(e) and (o) as well as Figs.~\ref{fig7}(h) and (k), and \ref{fig8}(c)-(e)
were obtained only via the problem-specific augmentation
described in section \ref{sec:intro},
rather than being discovered through deflation from other
branches. This serves as an important caveat: deflated continuation
should not be thought of as a universal solvent that blindly reveals all solutions
to a particular bifurcation problem, but as a useful tool that becomes
even more powerful when combined with physical insight.
In particular, a deep understanding of the underlying physics of the
system (in this case knowledge of the linear limit) remains
of paramount value in uncovering the complexity of its landscape
of stationary solutions.
It is conceivable that the robustness of deflated continuation could
be improved by reducing the continuation step-size or by employing
more robustly globalized nonlinear solvers,
thus enabling the method to detect more branches.
This issue merits further computational investigation.


Nevertheless, the significant
success of deflation continuation in the present setting
suggests that it is well suited to identifying steady
states in multi-dimensional PDEs with such energy
landscapes, providing insights on the connections between different
extrema (and saddle points) in them.
In the context of BECs there exists a wide array of problems
that are very much worth pursuing. A natural
extension is to attempt to generalize the methodology to
multi-component BECs~\cite{revip,book_new}.  Another
important extension is to three-dimensional configurations
in a single component, including dark solitons, vortex
rings, vortex lines, Hopfions
etc.~\cite{brand,ticknor1,ticknor2}. A further
generalization would be to three-dimensional multi-component
settings, where structures such as skyrmions and merons
arise~\cite{ruost1,ruost2,bigel}. Such studies are
currently in progress and will be reported in future
publications.

\begin{acknowledgments}
E.G.C and P.G.K. thank Ricardo Carretero (SDSU) for multiple fruitful 
discussions on this project. P.E.F.~acknowledges support from 
EPSRC grants EP/K030930/1 and EP/M019721/1, from a Center of 
Excellence grant from the Research Council of Norway to the Center for Biomedical
Computing at Simula Research Laboratory (project number
179578), and from the generous support of Sir Michael Moritz
and Harriet Heyman. P.G.K.~gratefully acknowledges support
from the Alexander von Humboldt Foundation,
the US-NSF under grants DMS-1312856, and PHY-1602994,
as well as the ERC under FP7, Marie
Curie Actions, People, International Research Staff Exchange
Scheme (IRSES-605096).
\end{acknowledgments}

\end{document}